\documentstyle[12pt,epsf,axodraw]{article}
\makeatletter
\def\appendix{\par\clearpage
  \setcounter{section}{0}
  \setcounter{subsection}{0}
  \@addtoreset{equation}{section}
  \def\@sectname{Appendix~}
  \def\theequation{\thesection.\arabic{equation}}
  \def\thesection{\Alph{section}}}
\makeatother

\renewcommand{\theequation}{\thesection.\arabic{equation}}

\begin{document}
\begin{titlepage}
\hskip 11cm \vbox{
\hbox{BUDKERINP/99-61}
\hbox{UNICAL-TH 99/3}
\hbox{July 1999}}
\vskip 0.3cm
\centerline{\bf THE GLUON IMPACT FACTORS$^{~\ast}$}
\vskip 0.8cm
\centerline{  V.S. Fadin$^{a~\dagger}$, R. Fiore$^{b~\ddagger}$,
M.I. Kotsky$^{a~\dagger}$, A. Papa$^{b~\ddagger}$}
\vskip .3cm
\centerline{\sl $^{a}$ Budker Institute for Nuclear Physics and Novosibirsk}
\centerline{\sl State University, 630090 Novosibirsk, Russia}
\centerline{\sl $^{b}$ Dipartimento di Fisica, Universit\`a della Calabria
and Istituto}
\centerline{\sl Nazionale di Fisica Nucleare, Gruppo collegato di Cosenza,}
\centerline{\sl Arcavacata di Rende, I-87036 Cosenza, Italy}
\vskip 1cm
\begin{abstract}
We calculate in the next-to-leading approximation the non-forward gluon 
impact factors for arbitrary color state in the $t$-channel. In the 
case of the octet state we check the so-called "second bootstrap condition" 
for the gluon Reggeization in QCD, using the integral representation for 
the impact factors. The condition is fulfilled in the general case of an
arbitrary space-time dimension and massive quark flavors for both 
helicity conserving and non-conserving parts.
\end{abstract}
\vfill
\hrule
\vskip.3cm
\noindent
$^{\ast}${\it Work supported in part by the Ministero italiano
dell'Universit\`a e della Ricerca Scientifica e Tecnologica, in part
by INTAS and in part by the Russian Fund of Basic Researches.}
\vfill
$ \begin{array}{ll}
^{\dagger}\mbox{{\it e-mail address:}} &
 \mbox{FADIN, KOTSKY ~@INP.NSK.SU}\\
\end{array}
$

$ \begin{array}{ll}
^{\ddagger}\mbox{{\it e-mail address:}} &
  \mbox{FIORE, PAPA ~@FIS.UNICAL.IT}
\end{array}
$
\vfill
\vskip .1cm
\vfill
\end{titlepage}
\eject

\section{Introduction}
\setcounter{equation}{0}

The BFKL equation~\cite{1} is widely discussed now, because it can 
enlighten an important question of elementary particle physics such as the 
theoretical description of QCD semi-hard processes. It gets a special 
importance due to the 
present experimental investigation of the deep inelastic electron-proton
scattering at HERA (see, for example,~\cite{2}) in the   region 
of small values of
the Bjorken variable $x$.
This equation was derived more than twenty years ago in the leading
logarithmic approximation
(LLA)~\cite{1}, where all the terms of the type $\alpha_s^n\ln^n(1/x)$ are 
summed up. Recently, the radiative  corrections to the equation were 
calculated~\cite{3}-\cite{8} and the explicit form of the kernel of the equation 
in the next-to-leading approximation (NLA) became known~\cite{9,CC98} 
for the case of forward scattering. The large size of the corrections 
induced a number of subsequent publications (see, for instance,~\cite{NLA}). 

In the BFKL approach the high energy scattering amplitudes are given as the 
convolution (see Eq.(\ref{Ar}) below) of the Green function for  two interacting 
Reggeized gluons with the impact factors of the colliding 
particles~\cite{1,9,10,11,12}. 
While the Green function is determined by the kernel of the  BFKL equation,  
the impact factors must be evaluated separately. In some cases such as,
for instance, the impact factors of strongly-virtual photons or hard mesons,
this can be done in perturbation theory, while in the general non-perturbative 
case there is need of new ideas for the evaluation.

This paper is devoted to the calculation of the NLA non-forward gluon impact
factors with arbitrary color state in the $t$-channel. Although they are not 
directly connected with observable cross sections,  their knowledge is
very important for the BFKL theory for two reasons. Firstly, these impact
factors can be used for the NLA calculation of scattering amplitudes of
partons in the BFKL approach. The second reason, on which we mainly concentrate
here, is the necessity to check the so-called "bootstrap" conditions~\cite{12}.
The matter is the following: the base of the BFKL equation approach is the 
property of the ``gluon Reggeization'', whose exact meaning was explained in 
details in Ref.~\cite{12}. This property was proved only in the 
LLA~\cite{13}, while beyond the LLA it was checked only in the first three 
orders of the perturbation theory. The "bootstrap" conditions, obtained in 
Ref.~\cite{12}, are just appealed to demonstrate the self-consistency of the 
BFKL approach in the NLA, although (if satisfied) they cannot be considered 
a proof of the Reggeization in a mathematical sense. The fulfillment of these 
conditions, however, would confirm so strongly the Reggeization 
that there would be no
doubts that it is correct. Moreover, the check of the "bootstrap" equations
is extremely important since they involve almost all the values appearing in 
the NLA BFKL kernel, so that the check provides a global test 
of the calculations~\cite{3}-\cite{8} of the NLA corrections, which only in a small 
part were independently performed~\cite{8} or checked~\cite{BRN98,DS98}. 

The first bootstrap equation derived in~\cite{12} connects the kernel of 
the non-forward BFKL equation for the color octet in the $t$-channel with 
the gluon trajectory. In Ref.~\cite{FFP98} it was shown that this equation 
is satisfied in the part concerning the quark contribution for arbitrary 
space-time dimension. The second bootstrap condition involves the 
impact factors of the scattered particles with color octet in the $t$-channel.
The case of colliding gluons is the object of the present paper, while
quarks have been considered in a related paper~\cite{FFKP99}.

The paper is organized as follows. In the next Section we explain the method of
calculation, Sections~3, 4 and 5 are devoted to the calculation of one-gluon,
quark-antiquark and two-gluon contributions to the gluon impact factors,
respectively, for arbitrary color group representation in the $t$-channel. 
Section~6 contains details of the check of the second bootstrap
condition, which involves the octet gluon impact factors in the NLA. The 
NLA gluon impact factors for the case of QCD with massless quark flavors are
considered in the Section~7. The results obtained are briefly discussed in
Section~8. Some integrations are carried out in the Appendix~A.

\section{Method of calculation}
\setcounter{equation}{0}

Let us remind that the
impact factors were introduced in the BFKL
approach for the description of the elastic scattering amplitudes
$A + B \rightarrow A^{\prime} + B^{\prime}$ in the Regge kinematical region
\begin{equation}\label{21}
s = ({p}_A + {p}_B)^2 = ({p}^{\prime}_A + 
{p}^{\prime}_B)^2 \rightarrow \infty,\ \ \ \ t = ({p}_A -
{p}^{\prime}_A)^2 = ({p}^{\prime}_B - {p}_B)^2\ \ \ \ 
\mbox{--fixed},
\end{equation}
where ${p}_A,~{p}_B$ and ${p}^{\prime}_A,
p^{\prime}_B$ are the momenta of the initial and final particles,
respectively. We use for all vectors the Sudakov decomposition
\begin{equation}\label{22}
p = \beta p_1 + \alpha p_2 + p_{\perp},\ \ \ \ \ \ \ \ p_{\perp}^2 = 
- \vec p^{\:2}~,
\end{equation}
the vectors $(p_1,\ p_2)$ being the light-cone basis of the initial
particle momenta plane $(p_A,\ p_B)$, so that we can put
\begin{equation}\label{23}
p_A= p_1 +
\frac{m_A^2}{s}p_2~,\ \ \ \ \ \ \ p_B =p_2 +
\frac{m_B^2}{s} p_1~.
\end{equation}
Here $m_A$ and $m_B$ are the masses of the colliding particles $A$ and $B$ and
the vector notation is used throughout this paper for the transverse
components of the momenta, since all vectors in the transverse subspace are 
evidently space-like.

The basis of the BFKL approach is the gluon Reggeization. In the case of 
the elastic scattering, it means that the amplitude with gluon quantum 
numbers and negative signature in the $t$-channel has the Regge form
\begin{equation}
({\cal A}_{8}^{(-)})_{AB}^{A^{\prime }B^{\prime }}=\Gamma _{A^{\prime }A}
^{c}\left[ \left( \frac{-s}{-t}\right) ^{j(t)}-\left( \frac{+s}{-t}\right)
^{j(t)} \right] \Gamma _{B^{\prime }B\mbox{ \ \ }}^{c} .  
\label{Ao}
\end{equation}
Here $c$ is a color index, $\Gamma _{P^{\prime }P}^{c}$ are the 
particle-particle-Reggeon (PPR) vertices which do not depend on $s$ and
$j(t) = 1 + \omega(t)$ is the Reggeized gluon trajectory.
In the derivation of the BFKL equation in the NLA it was assumed that 
this form, as well as the multi-Regge form of production amplitudes (see, 
for instance,~\cite{12} and references therein) is  
valid also in the NLA. Then, the $s$-channel unitarity of the scattering
matrix leads to

\begin{figure}[tb]      
\begin{center}
\begin{picture}(240,200)(0,0)

\ArrowLine(0,190)(75,190)
\ArrowLine(165,190)(240,190)
\Text(37.5,200)[]{$p_A$}
\Text(202.5,200)[]{$p_{A'}$}
\Text(120,190)[]{$\Phi_{A'A}(\vec q_1,\vec q)$}
\Oval(120,190)(20,45)(0)

\Gluon(100,135)(100,172){4}{3}
\Gluon(140,172)(140,135){4}{3}

\Gluon(100,28)(100,65){4}{3}
\Gluon(140,65)(140,28){4}{3}

\ArrowLine(96,158)(96,156)
\ArrowLine(144,156)(144,158)

\ArrowLine(96,44)(96,42)
\ArrowLine(144,42)(144,44)

\Text(85,157)[]{$q_1$}
\Text(165,157)[]{$q_1-q$}

\Text(85,43)[]{$q_2$}
\Text(165,43)[]{$q_2-q$}

\GCirc(120,100){40}{1}
\Text(120,100)[]{$G(\vec q_1,\vec q_2;\vec q)$}

\ArrowLine(0,10)(75,10)
\ArrowLine(165,10)(240,10)\Text(37.5,0)[]{$p_B$}
\Text(202.5,0)[]{$p_{B'}$}
\Text(120,10)[]{$\Phi_{B'B}(-\vec q_2,-\vec q)$}
\Oval(120,10)(20,45)(0)

\end{picture}
\end{center}
\caption[]{Diagrammatic representation of the elastic scattering amplitude
$A + B \rightarrow A' + B'$.} 
\end{figure}
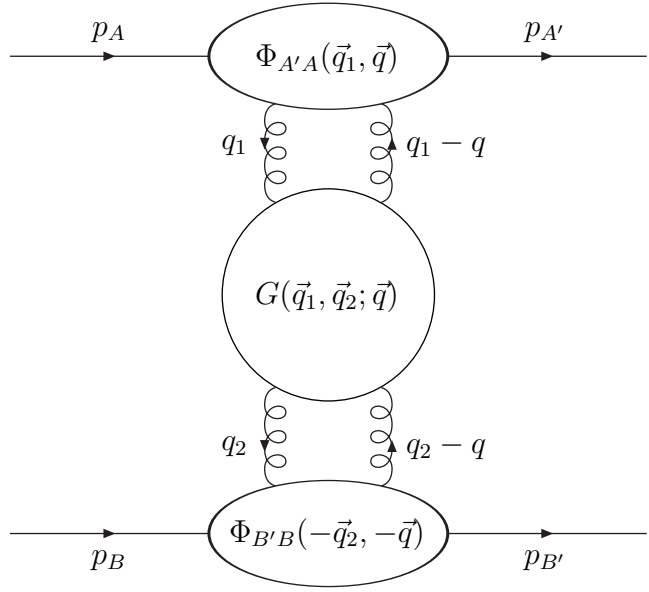

\[
{\cal I}{\it m}_{s}\left(({\cal A}_{{\cal R}})_{AB}^{A^{\prime}
B^{\prime }}\right) = \frac{s}{\left( 2\pi \right)^{D-2}}
\int \frac{d^{D-2}q_1}{\vec{q}_{1}^{\:2} \vec{q}_{1}^{\:\prime\:2}}
\int \frac{d^{D-2}q_2}{\vec{q}_{2}^{\:2} \vec{q}_{2}^{\:\prime\:2}}
\]
\begin{equation}
\times \sum_{\nu}\Phi _{A^\prime A}^{\left( {\cal R},\nu \right) }
\left( \vec{q}_{1},\vec{q};s_{0}\right)\int_{\delta -i\infty}^{\delta+i\infty} 
\frac{d\omega }{2\pi i}\left[ \left( \frac{s}{s_{0}}\right)^{\omega }
G_{\omega }^{\left( {\cal R}\right) }\left( \vec{q}_{1},\vec{q}_{2},\vec{q}
\right) 
\right] \Phi _{B^\prime B}^{\left( {\cal R},\nu \right) }\left( -\vec{q}_{2},
-\vec{q};s_{0}\right) \;,
\label{Ar}
\end{equation}
where the momenta are defined in Fig.~1. For convenience we have introduced  
the notation (which will be used also in the 
following) $q_i^\prime \equiv q_i - q$, where $q\simeq q_{\perp}$ is the 
momentum transfer in the process $A + B \rightarrow A^{\prime} 
+ B^{\prime}$ .  We emphasize that the wavy intermediate lines    
in Fig.~1 denote Reggeons and not gluons -- the Reggeons would only become
gluons in absence of interaction. The space-time dimension, $D$, is    
taken to be $D = 4 + 2 \epsilon$ in order to regularize the infrared    
divergences.    
In the above equation ${\cal A}_{\cal R}$ stands for the scattering amplitude with
the irreducible representation ${\cal R}$ of the color group in the 
$t$-channel,
the index $\nu$ enumerates the states in this representation, 
$\Phi_{P^\prime P}^{\left( {\cal R},\nu \right)}$
are the impact factors and $G_{\omega }^{\left( {\cal R}\right)}$ is the Mellin 
transform of the Green function for the Reggeon-Reggeon 
scattering~\cite{12}.
Here and below we do not indicate the signature, since it is defined by the symmetry
of the representation ${\cal R}$ in the product of the two octet representations.
The parameter $s_0$ is an arbitrary energy scale introduced in order to define 
the partial wave expansion of the scattering amplitudes.  The dependence 
on this parameter disappears in the full expressions 
for the amplitudes. The Green function obeys the generalized BFKL equation
\begin{equation}
\omega G_{\omega }^{\left( {\cal R}\right) }\left( \vec{q}_{1},\vec{q}_{2},
\vec{q}\right) = \vec{q}_{1}^{\:2} \vec{q}_{1}^{\:\prime\:2}
\delta^{\left(D-2\right) }\left( \vec{q}_{1}-\vec{q}_{2}\right)
+\int \frac{d^{D-2} q_r }
{\vec{q}_r^{\:2} \vec{q}_r^{\:\prime \:2}}{\cal K}
^{\left( {\cal R}\right) }\left( \vec{q}_{1},\vec{q}_r,\vec{q}
\right) G_{\omega }^{\left( {\cal R}\right) }\left( \vec{q}_r,\vec{q}
_{2},\vec{q} \right) \;,
\label{genBFKL}
\end{equation}
where ${\cal K}^{\left( {\cal R}\right)}$ is the kernel in the
NLA~\cite{12}.

\begin{figure}
\begin{center}
{\parbox[t]{5cm}{\epsfysize 5cm \epsffile{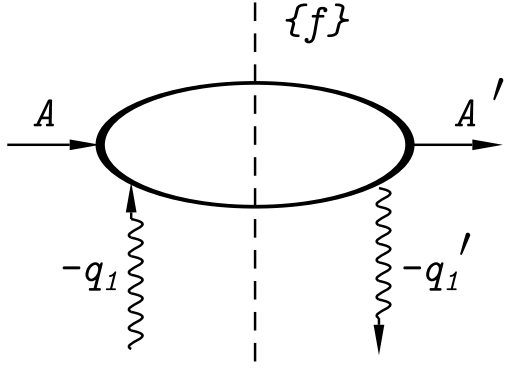}}}
\end{center}
\caption{Schematic description of the intermediate states contributions
to the impact factors.}
\label{fig1}
\end{figure}

The bootstrap conditions appear from the requirement that the imaginary 
part of the amplitude (\ref{Ao}) must coincide with the R.H.S. of Eq. 
(\ref{Ar}) in the case
of gluon quantum numbers in the $t$-channel.
The second bootstrap condition in the NLA reads~\cite{12}
$$
- \int\frac{d^{D-2}q_1}{(2\pi)^{D-1}}\frac{\vec q^{\:2}}{\vec q_1^{\:2}
\vec{q}_{1}^{\:\prime\:2}}ig\sqrt{N}\Phi_{A^{\prime}A}^{(8,a)(1)}
(\vec q_1, \vec q;s_0) =
$$
\begin{equation}\label{24}
\Gamma_{A^{\prime}A}^{(a)(1)}\omega^{(1)}( - \vec q^{\:2})
+ \frac{1}{2}\Gamma_{A^{\prime}A}^{(a)(B)}\left[ \omega^{(2)}( - \vec
q^{\:2}) + \left( \omega^{(1)}( - \vec q^{\:2}) \right)^2\ln\left( \frac{s_0}
{\vec q^{\:2}} \right) \right]~.
\end{equation}
Here  $g$ is the gauge coupling constant
($g^2 = 4\pi\alpha_s$), $N$ is the number of colors,
$\omega^{(1)}$ and $\omega^{(2)}$ are the one- and two-loop contributions
to the Reggeized gluon trajectory, $\Gamma_{A^{\prime}A}^{(a)(B)}$ and  
$\Gamma_{A^{\prime}A}^{(a)(1)}$ are  
the Born and one-loop parts of the particle-particle-Reggeon (PPR) effective
vertex.
The definition of the non-forward impact factors with color state $\nu$ of 
the irreducible representation ${\cal R}$ was given in Ref.~\cite{12} and can 
be presented as
\begin{equation}\label{27}
\Phi_{A^{\prime}A}^{({\cal R}, \nu)}(\vec q_1, \vec q; s_0) = \langle
cc^{\prime} | \hat{\cal P}_{\cal R} | \nu \rangle\Phi_{AA^{\prime}}^
{cc^{\prime}}(\vec q_1, \vec q; s_0)~,
\end{equation}
where $\hat{\cal P}_{\cal R}$ is the projector of two-gluon color states
in the $t$-channel on the irreducible representation ${\cal R}$. 
The value $\Phi_{AA^{\prime}}^{cc^{\prime}}$ determines
completely the impact factors of the particle $A$ with any possible color
structure. We will consider mainly just this object and call it  
unprojected impact factor. The definition of such impact factors 
in the NLA can be
reconstructed from the one of Ref.~\cite{12}:
$$
\Phi_{AA^{\prime}}^{cc^{\prime}}(\vec q_1, \vec q; s_0) = \left( \frac{s_0}
{\vec q_1^{\:2}} \right)^{\frac{1}{2}\omega( - \vec q_1^{\:2})}
\left( \frac{s_0}{\vec q_1^{\:\prime\:2}} \right)^{\frac{1}{2}
\omega( - \vec q_1^{\:\prime\:2})}\sum_{\{f\}}\int\theta(s_{\Lambda} -
s_{AR})\frac{ds_{AR}\: d\rho_f}{(2\pi)}\Gamma_{\{f\}A}^c
$$
\begin{equation}\label{28}
\times\left( \Gamma_{\{f\}A^{\prime}}^{c^{\prime}} \right)^* -
\frac{1}{2}\int\frac{d^{D-2}q_2}{\vec q_2^{\:2}\vec{q}_{2}^{\:\prime\:2}}
\: \Phi_{AA^{\prime}}^{c_1c_1^{\prime}(B)}(\vec q_2, \vec q)
\: ({\cal K}^B_r)_{c_1c}^{c_1^{\prime}c^{\prime}}
(\vec q_2, \vec q_1, \vec q)\:\ln\left(\frac{s_{\Lambda}^2}
{s_0(\vec q_2 - \vec q_1)^2} \right)~.
\end{equation}
In this expression it is enough to take  the Reggeized 
gluon trajectory $\omega(t)$ in the one-loop approximation. For brevity, 
we do not perform here and below an explicit expansion in $\alpha_s$; evidently,
this expansion is assumed and only the leading and the next-to-leading 
terms should be kept.
$\Gamma_{\{f\}A}^c$ is the effective vertex for production of the
system ${\{f\}}$ (see Fig.~\ref{fig1}) in the collision of the particle $A$
and the Reggeized gluon with color index $c$ and momentum
\begin{equation}\label{29}
- q_1 = \alpha p_2 - {q_1}_{\perp}~,\ \ \ \ \ \ \ \alpha 
\approx \left( s_{AR} - m_A^2 + \vec q_1^{\:2} \right)/s \ll 1~, 
\end{equation}
and $s_{AR}$ is the particle-Reggeon squared invariant mass. 
In the fragmentation region of the particles $A$ and $A^{\prime}$, 
where all transverse momenta as well as the invariant mass $\sqrt{s_{AR}}$ are not 
growing with $s$, we have for both Reggeons the relations
\begin{equation}\label{212}
q_1^2 = - \vec q_1^{\:2}~,\ \ \ \ \ \ q_1^{\:\prime\:2} = - \vec q_1^{\:\prime\:2} =
- (\vec q_1 - \vec q)^2.
\end{equation}
Summation in
Eq.~(\ref{28}) is carried out over all  systems $\{f\}$ which can be
produced in the NLA and the integration is performed over the phase
space volume of the produced system, which for a $n$-particle
system (if there are identical particles in this system, 
corresponding factors should also be introduced) reads
\begin{equation}\label{213}
d\rho_f = (2\pi)^D\delta^{(D)}\biggl(p_A - q_1 - \sum_{m=1}^nk_m\biggr)
\prod_{m=1}^n\frac{d^{D-1}k_m}{2\epsilon_m(2\pi)^{D-1}}~,
\end{equation}
as well as over the particle-Reggeon invariant mass.
The parameter $s_{\Lambda}$, limiting the integration region over the invariant
mass in the first term in the R.H.S. of Eq.~(\ref{28}), is introduced 
for the separation of the contributions of multi-Regge and quasi-multi-Regge
kinematics (MRK and QMRK) and should be considered as tending to infinity.
The dependence of the impact factors on this parameter disappears~\cite{12} 
due to the cancellation between the first and the second term in
the R.H.S. of Eq.~(\ref{28}). 
In the second term, $\Phi_{AA^{\prime}}^{cc^{\prime}(B)}$ is the Born
contribution to the impact factor, which does not depend on $s_0$ (for this reason
we omitted this argument there), while
$({\cal K}^B_r)_{c_1c}^{c_1^{\prime}c^{\prime}}$ is the part of the  
unprojected non-forward BFKL kernel in the Born approximation 
connected with real particle production:
\begin{equation}\label{214}
({\cal K}^B_r)_{c_1c}^{c_1^{\prime}c^{\prime}}(\vec q_1, \vec q_2, \vec q) =
\frac{g^2}{(2\pi)^{D-1}}T^d_{c_1c}T^d_{c^{\prime}c_1^{\prime}}\left(
\frac{\vec q_1^{\:2}\vec q_2^{\:\prime\:2} + \vec q_2^{\:2}\vec q_1^{\:\prime\:2}}
{(\vec q_1 - \vec q_2)^2} - \vec q^{\:2} \right),
\end{equation}
being $T$ the color group generator in the adjoint representation.
Let us note that the definitions in 
Eqs.~(\ref{27}) and (\ref{28}) imply a suitable normalization of the amplitudes
$\Gamma_{\{f\}A}^c$, namely that applied in Eq.~(11) of 
Ref.~\cite{12} (see also there the text
after Eq.~(27)). We should also note, that the definitions~(\ref{27}) and 
(\ref{28}) are applicable for the case of colorless particles as
well as for the case of charged QCD particles, while the octet impact
factors 
$\Phi_{A^{\prime}A}^{(8,a)}$, entering the bootstrap condition (\ref{24}),
have sense, of course, only for colored particles.

Considering the impact factors of the particle A, we can without loss of generality 
assume the particle $B$ to be massless, because 
the impact factors $\Phi_{AA^{\prime}}^{cc^{\prime}}$ are properties of the
particle $A$ only and cannot depend on the properties of the other scattered
particle. So, everywhere below the initial particle momenta
$p_A$ and $p_B$ are taken as the light-cone basis. For any gluon polarization vector
we will use the light-cone gauge
\begin{equation}\label{215}
e(k)\: p_B = 0 
\end{equation}
and from the transversality of this vector to the gluon momentum $k$ it is
easy to get the following Sudakov representation:
\begin{equation}\label{216}
e(k) = - \frac{\left( k_{\perp}e_{\perp}(k) \right)}{(kp_B)}\:p_B + 
e_{\perp}(k)~.
\end{equation}
The transverse polarization vectors have the properties
$$
\left( e_{\perp}^*(k, \lambda_1)e_{\perp}(k, \lambda_2) \right) =
\left( e^*(k, \lambda_1)e(k, \lambda_2) \right) = - \delta_{\lambda_1,\,
\lambda_2},
$$
\begin{equation}\label{217}
\sum_{\lambda}e_{\perp}^{*\mu}(k, \lambda)e_{\perp}^{\nu}(k, \lambda) =
- g_{\perp\perp}^{\mu\nu}~,
\end{equation}
where the index $\lambda$ enumerates the independent polarizations of gluon,
$g^{\mu\nu}$ is the metrical tensor in the full space and 
$g_{\perp\perp}^{\mu\nu}$ the one in the transverse subspace, 
\begin{equation}\label{218}
g_{\perp\perp}^{\mu\nu} = g^{\mu\nu} - 
\frac{p_A^{\mu}p_B^{\nu} + p_B^{\mu}p_A^{\nu}}
{(p_Ap_B)}~.
\end{equation}
With the NLA accuracy the intermediate states $\{f\}$, which can contribute
to the impact factors (\ref{28}) in the gluon case, are one-gluon, 
two-gluon
and quark-antiquark-pair states. In the case of the two-gluon contribution, we 
include the second term in the R.H.S. of Eq.~(\ref{28}), which is a counterterm 
for the LLA part contained in the first term. In the Born and
quark-antiquark contributions to the gluon impact factors we will omit the
argument $s_0$, because of their evident independence on it. Firstly, we will
consider the general case of
arbitrary $\epsilon = (D-4)/2$ and arbitrary mass $m_f$ for the quark flavor $f$.
Under these general conditions, we will determine integral representations for
the gluon impact factors. These general integral representations are necessary 
for the check of the bootstrap condition (\ref{24}). Then we will perform the 
integration in the expansion in $\epsilon$ for the practically important case of
QCD with $n_f$ massless quark flavors. 

\section{One-gluon contribution}
\setcounter{equation}{0}

In the case of the one-gluon contribution, the invariant mass $\sqrt{s_{AR}}$ is 
fixed to be zero, because of the masslessness of the intermediate gluon $G$, and one
easily gets from the definition (\ref{28})
\begin{equation}\label{31}
\Phi_{AA^{\prime}}^{cc^{\prime}\{G\}}(\vec q_1, \vec q; s_0) = \left(
\frac{s_0}{\vec q_1^{\:2}} \right)^{\frac{1}{2}\omega( - \vec q_1^{\:2})}
\left(\frac{s_0}{\vec q_1^{\:\prime\:2}} \right)^{\frac{1}{2}
\omega( - \vec q_1^{\:\prime\:2})}\sum_{\lambda}\Gamma^c_{GA}
\left( \Gamma^{c^{\prime}}_{GA^{\prime}} \right)^*~,
\end{equation}
where  the gluon-gluon-Reggeon (GGR) effective vertex
$\Gamma^c_{GA}$, obtained in Refs.~\cite{6,14} and~\cite{15}, has the 
form 
\begin{equation}\label{32}
\Gamma^c_{GA} = gT^c_{GA}\left[ \delta_{\lambda, \lambda_A}\left(
1+ \Gamma_{GG}^{(+)(1)}(q_1^2) \right) + \delta_{\lambda, - \lambda_A}
\Gamma_{GG}^{(-)(1)}(q_1^2) \right]~.
\end{equation}
Here $\Gamma_{GG}^{(\pm)(1)}$ represent the radiative corrections to 
the helicity con\-ser\-ving and non-con\-ser\-ving parts of the vertex,
the $\delta$'s on the helicities
$\lambda_A$ and $\lambda$ of the gluons $A$ and $G$, respectively, are
determined by their form in the tensor representation
$$
\delta_{\lambda, \pm\lambda_A} = e_{\mu}^*e_{A\mu_A}\delta_{\lambda,
\pm\lambda_A}^{\mu\mu_A},\ \ \delta_{\lambda, \lambda_A}^{\mu\mu_A} =
- \left( g^{\mu_A\rho} - \frac{p_A^{\mu_A}p_B^{\rho} + p_B^{\mu_A}p_A^{\rho}}
{(p_Ap_B)} \right)\left( g^{\mu\nu} - \frac{p^{\mu}p_B^{\nu} + p_B^{\mu}
p^{\nu}}{(pp_B)} \right)g_{\rho\nu}~,
$$
\begin{equation}\label{33}
\delta_{\lambda, - \lambda_A}^{\mu\mu_A} = - \left( g^{\mu_A\rho} - \frac
{p_A^{\mu_A}p_B^{\rho} + p_B^{\mu_A}p_A^{\rho}}{(p_Ap_B)} \right)\left(
g^{\mu\nu} - \frac{p^{\mu}p_B^{\nu} + p_B^{\mu}p^{\nu}}{(pp_B)} \right)\left(
g_{\rho\nu} - (D - 2)\frac{q_{1\rho}q_{1\nu}}{q_1^2} \right)~,
\end{equation}
where $p$ is the momentum of the intermediate gluon $G$. Using the 
Eqs.~(\ref{216}) and (\ref{33}), it is easy to obtain
the following relations:
$$
\delta_{\lambda, \lambda_A} = - e_{\perp}^{*\mu}e_{A\perp}^{\nu}g_{\mu\nu}
^{\perp\perp}~,\ \ \ \ \ \ \delta_{\lambda, - \lambda_A} =
- e_{\perp}^{*\mu}e_{A\perp}^{\nu}T_{\mu\nu}^{\perp\perp}(q_{1\perp})~,
$$
\begin{equation}\label{34}
\delta_{\lambda, \lambda_{A^{\prime}}} = - e_{\perp}^{*\mu}e_{A^{\prime}\perp}
^{\nu}g_{\mu\nu}^{\perp\perp}~,\ \ \ \ \ \
\delta_{\lambda, - \lambda_{A^{\prime}}} = 
-e_{\perp}^{*\mu}e_{A^{\prime}\perp}
^{\nu}T_{\mu\nu}^{\perp\perp}(q_{1\perp}^{\prime})~,
\end{equation}
with
\begin{equation}\label{35}
T_{\mu\nu}^{\perp\perp}(k_{\perp}) = g_{\mu\nu}^{\perp\perp} - (D - 2)\frac
{k_{\mu}^{\perp}k_{\nu}^{\perp}}{k_{\perp}^2}~.
\end{equation}
From these relations one can get without any difficulty
$$
\sum_{\lambda}\delta_{\lambda, \lambda_A}\delta_{\lambda, \lambda_{A^{\prime}}}
^* = - e_{A^{\prime}\perp}^{*\mu}e_{A\perp}^{\nu}g_{\mu\nu}^{\perp\perp}~,
$$
$$
\sum_{\lambda}\delta_{\lambda, - \lambda_A}\delta_{\lambda, \lambda_{A^
{\prime}}}^* = - e_{A^{\prime}\perp}^{*\mu}e_{A\perp}^{\nu}T_{\mu\nu}^
{\perp\perp}(q_{1\perp})~,
$$
\begin{equation}\label{36}
\sum_{\lambda}\delta_{\lambda, \lambda_A}\delta_{\lambda, - \lambda_
{A^{\prime}}}^* = - e_{A^{\prime}\perp}^{*\mu}e_{A\perp}^{\nu}T_{\mu\nu}^
{\perp\perp}(q_{1\perp}^{\prime})~,
\end{equation}
which gives for the convolution in Eq.~(\ref{31}):
$$
\Phi_{AA^{\prime}}^{cc^{\prime}\{G\}}(\vec q_1, \vec q; s_0) = - g^2
\left( T^{c^{\prime}}T^c \right)_{A^{\prime}A}e_{A^{\prime}\perp}^{*\mu}
e_{A\perp}^{\nu}\biggl[ g_{\mu\nu}^{\perp\perp}\biggl( 1 + \Gamma_{GG}^{(+)(1)}
(- \vec q_1^{\:2})
$$
$$
+ \Gamma_{GG}^{(+)(1)}(- \vec q_1^{\:\prime\:2}) + \frac{1}{2}\omega^{(1)}
(- \vec q_1^{\:2})\ln\left( \frac{s_0}{\vec q_1^{\:2}} \right) + \frac{1}{2}
\omega^{(1)}(- \vec q_1^{\:\prime\:2})\ln\left( \frac{s_0}{\vec q_1^{\:\prime\:2}}
\right) \biggr)
$$
\begin{equation}\label{37}
+ T_{\mu\nu}^{\perp\perp}(q_{1\perp})\Gamma_{GG}^{(-)(1)}
(- \vec q_1^{\:2}) + T_{\mu\nu}^{\perp\perp}(q_{1\perp}^{\prime})
\Gamma_{GG}^{(-)(1)}(- \vec q_1^{\:\prime\:2}) \biggr]~.
\end{equation}

We need now the expressions for the radiative corrections 
$\Gamma_{GG}^{(\pm)(1)}$, which can be found in Ref.~\cite{14}. We present 
these expressions in a form slightly different from the one used there:
$$
\Gamma_{GG}^{(+)(1)}(- \vec v^{\:2}) = g^2N\frac{ \Gamma(1 -
\epsilon)}{(4\pi)^{2 + \epsilon}}\frac{\Gamma^2(1 + \epsilon)}{\Gamma(1 +
2\epsilon)}\frac{1}{\epsilon}\biggl( - \frac{2}{\epsilon} + \frac{9(1 +
\epsilon)^2 + 2}{2(1 + \epsilon)(1 + 2\epsilon)(3 + 2\epsilon)}
$$
$$
+ 2\psi(1 + \epsilon) - \psi(1 - \epsilon) - \psi(1) \biggr)(\vec
v^{\:2} )^{\epsilon} + g^2\frac{\Gamma(1 -
\epsilon)}{(1 + \epsilon)(4\pi)^{2 + \epsilon}}
$$
$$
\times\biggl[ \sum_f\frac{2(1 + \epsilon)}{\epsilon}\int_0^1dxx(1 - x)\left(
m_f^2 + x(1 - x)\vec v^{\:2} \right)^{\epsilon} - \frac{1}{2}
F_1^{(+)}(\vec v^{\:2}) \biggr],
$$
$$
\Gamma_{GG}^{(-)(1)}(- \vec v^{\:2}) = g^2N\frac{\Gamma(1 -
\epsilon)}{(4\pi)^{2 + \epsilon}}\frac{\Gamma^2(1 + \epsilon)}{\Gamma(1 +
2\epsilon)}
$$
\begin{equation}\label{38}
\times\frac{1}{(1 + \epsilon)(1 + 2\epsilon)(3 + 2\epsilon)}\left(
\vec v^{\:2} \right)^{\epsilon} - g^2\frac{\Gamma
(1 - \epsilon)}{(1 + \epsilon)(4\pi)^{2 + \epsilon}}2F_1^{(-)}(\vec v^{\:2})~,
\end{equation}
with
$$
F_1^{(+)}(\vec v^{\:2}) = \sum_f\biggl[ \frac{2}{3}\left(m_f^2
\right)^{\epsilon} + \int_0^1\int_0^1dx_1dx_2\theta(1 - x_1 - x_2)\left(
m_f^2 + x_1x_2\vec v^{\:2} \right)^{\epsilon}
$$
$$
\times\frac{\vec v^{\:2}(x_1 + x_2)\left( 1 + \epsilon -2(x_1 + x_2)(1 - x_1
- x_2) \right) - 4m_f^2(1 -x_1 -x_2)}{(m_f^2 + x_1x_2\vec v^{\:2})} \biggr]~,
$$
\begin{equation}\label{39}
F_1^{(-)}(\vec v^{\:2}) = \sum_f\int_0^1\int_0^1dx_1dx_2\theta(1 - x_1 - x_2)
\left( m_f^2 + x_1x_2\vec v^{\:2} \right)^{\epsilon}\frac
{\vec v^{\:2}x_1x_2(1 -x_1 -x_2)}{(m_f^2 + x_1x_2\vec v^{\:2})}~.
\end{equation}
In these equations $\Gamma(x)$ and $\psi(x)$ 
are the Euler gamma-function and its logarithmic derivative,
respectively. In order to pass from the representation of Ref.~\cite{14}
for the radiative corrections $\Gamma_{GG}^{(\pm)(1)}$ to our expressions,
we have used the identity
$$
0 = \frac{1}{2}F_1^{(+)}(\vec v^{\:2}) + \sum_f\biggl[ \frac{2(1 + \epsilon)}
{3\epsilon}\left( m_f^2 \right)^{\epsilon} - \int_0^1\int_0^1
dx_1dx_2\theta(1 - x_1 - x_2)\left( m_f^2 + x_1x_2\vec v^{\:2}
\right)^{\epsilon}
$$
\begin{equation}\label{310}
\times\frac{1}{(m_f^2 + x_1x_2\vec v^{\:2})}\biggl( \frac{(1 + \epsilon)}
{\epsilon}(2 - x_1 - x_2)\left( m_f^2 + (1 + 2\epsilon)\vec v^{\:2}x_1x_2
\right) - 2\vec v^{\:2}x_1x_2(1 - x_1 - x_2) \biggr) \biggr]~,
\end{equation}
that can be found also in Ref.~\cite{14}. We need also the expression
for the one-loop Reggeized gluon trajectory $\omega^{(1)}$~\cite{1}
\begin{equation}\label{311}
\omega^{(1)}(- \vec v^{\:2}) = - \frac{g^2N}{2}\int\frac{d^{D-2}k}{(2\pi)
^{D-1}}\frac{\vec v^{\:2}}{\vec k^{\:2}(\vec k - \vec v)^2} = - g^2N\frac
{\Gamma(1 - \epsilon)}{(4\pi)^{2 + \epsilon}}\frac{\Gamma^2(1
+ \epsilon)}{\Gamma(1 + 2\epsilon)}\frac{2}{\epsilon}\left( \vec
v^{\:2} \right)^{\epsilon}~.\ \ \ \
\end{equation}

Using Eqs.~(\ref{37})-(\ref{311}), we obtain finally the Born impact factor,
\begin{equation}\label{312}
\Phi_{AA^{\prime}}^{cc^{\prime}(B)}(\vec q_1, \vec q) = 
- g^2\left(T^{c^{\prime}}T^c \right)_{A^{\prime}A}e_{A^{\prime}\perp}^{*\mu}
e_{A\perp}^{\nu}g_{\mu\nu}^{\perp\perp},
\end{equation}
and the NLA correction to the impact factors due to the one-gluon
contribution, 
$$
\Phi_{AA^{\prime}}^{cc^{\prime}(1)\{G\}}(\vec q_1, \vec q; s_0) = - N\left(
T^{c^{\prime}}T^c \right)_{A^{\prime}A}g^4\frac{\Gamma(1 -
\epsilon)}{(4\pi)^{2 + \epsilon}}\frac{\Gamma^2(1 + \epsilon)}{\Gamma(1 +
2\epsilon)}e_{A^{\prime}\perp}^{*\mu}e_{A\perp}^{\nu}\biggl[ \biggl\{
g_{\mu\nu}^{\perp\perp}\frac{1}{\epsilon}
$$
$$
\times\biggl( - \ln\left( \frac{s_0}{\vec v^{\:2}} \right) - \frac{2}{\epsilon}
+ \frac{9(1 + \epsilon)^2 + 2}{2(1 + \epsilon)(1 + 2\epsilon)(3 + 2\epsilon)}
+ 2\psi(1 + \epsilon) - \psi(1 - \epsilon) - \psi(1) \biggr)\left( \vec
v^{\:2} \right)^{\epsilon}
$$
$$
+ \frac{1}{(1 + \epsilon)(1 + 2\epsilon)(3 + 2\epsilon)}\left( \vec
v^{\:2} \right)^{\epsilon}T_{\mu\nu}^{\perp\perp}(v_{\perp}) \biggr\}
\biggr|_{v_{\perp} = q_{1\perp}^{\prime}} + \biggl\{ \cdots \biggr\}\biggr|
_{v_{\perp} = q_{1\perp}} \biggr] - \left( T^{c^{\prime}}T^c \right)_
{A^{\prime}A}
$$
$$
\times g^4\frac{\Gamma(1 - \epsilon)}{(1 + \epsilon)(4\pi)^
{2 + \epsilon}}e_{A^{\prime}\perp}^{*\mu}e_{A\perp}^{\nu}\biggl[ \biggl\{
g_{\mu\nu}^{\perp\perp}\biggl( \sum_f\frac{2(1 + \epsilon)}{\epsilon}\int_0^1
dxx(1 - x)\left( m_f^2 + x(1 - x)\vec v^{\:2} \right)^{\epsilon}
$$
\begin{equation}\label{313}
- \frac{1}{2}F_1^{(+)}(\vec v^{\:2}) \biggr) - 2T_{\mu\nu}^{\perp\perp}
(v_{\perp})F_1^{(-)}(\vec v^{\:2}) \biggr\}\biggr|_{v_{\perp} = q_{1\perp}^
{\prime}} + \biggl\{ \cdots \biggr\}\biggr|_{v_{\perp} = q_{1\perp}} \biggr]~.
\end{equation}

\section{Quark-antiquark-pair contribution}
\setcounter{equation}{0}

In this section we calculate the pure NLA contribution to the gluon impact
factors defined in Eq.~(\ref{28}), coming from a quark-antiquark-pair
production in the fragmentation region:
\begin{equation}\label{41}
\Phi_{AA^{\prime}}^{cc^{\prime}(1)\{q\bar q\}}(\vec q_1, \vec q) = \sum_f\sum_
{\stackrel{\lambda_1, \lambda_2}{i_1, i_2}}\int\frac{ds_{AR}\:d\rho_
{\{q\bar q\}}}{(2\pi)}\Gamma^c_{\{q\bar q\}A}\left( \Gamma^{c^{\prime}}_
{\{q\bar q\}A^{\prime}} \right)^*~,
\end{equation}
where the summation is performed over the quark flavors $f$, over the helicities
$\lambda_1$ and $\lambda_2$ and over the color indices $i_1$ and $i_2$ of the
produced quark  and antiquark with momenta $k_1$ and $k_2$,
respectively. Integration over the invariant mass here is convergent
and we do not need to introduce the cutoff $s_{\Lambda}$. As usual, we use the
Sudakov representation
\begin{equation}\label{42}
k_{1, 2} = \beta_{1, 2}\:p_A + \frac{m_f^2 + \vec k_{1, 2}^{\:2}}{s\beta_{1, 2}}
p_B + k_{1, 2\perp}~,\ \ \ \ \ \ k_{1, 2}^2 = m_f^2~,
\end{equation}
where $m_f$ is the quark mass. Then we have the relations
\begin{equation}\label{43}
s_{AR} = (k_1 + k_2)^2 = \frac{m_f^2 + (\vec k_1\beta_2 - \vec k_2\beta_1)^2}
{\beta_1\beta_2}~,
\end{equation}
\begin{equation}\label{44}
\frac{ds_{AR}\:d\rho_{\{q\bar q\}}}{(2\pi)} = \delta(1 - \beta_1 - \beta_2)
\delta^{(D-2)}\left( (k_1 + k_2 + q_1)_{\perp} \right)\frac{d\beta_1d\beta_2}
{\beta_1\beta_2}\:\frac{d^{D-2}k_1 \, d^{D-2}k_2}{2(2\pi)^{D-1}}~.
\end{equation}
To obtain the last of these relations we have used also
Eqs.~(\ref{29}) and (\ref{213}).
The amplitude of $q\bar q$ pair production in the gluon-Reggeon collision
$\Gamma^c_{\{q\bar q\}A}$ was obtained in Ref.~\cite{14}, but we will use a 
more convenient, slightly different form for it, which can be easily obtained
from the expression presented there:
$$
\Gamma^c_{\{q\bar q\}A} = g^2\left[ \left( t^At^c \right)_{i_1i_2}
{\cal A}_{q\bar q} - \left( t^ct^A \right)_{i_1i_2}{\cal A}_{q\bar q}
(1 \leftrightarrow 2) \right]~,
$$
\begin{equation}\label{45}
\Gamma^{c^{\prime}}_{\{q\bar q\}A^{\prime}} = g^2\left[
\left( t^{A^{\prime}}t^{c^{\prime}} \right)_{i_1i_2}{\cal A}_{q\bar q}^{\prime}
- \left( t^{c^{\prime}}t^{A^{\prime}} \right)_{i_1i_2}{\cal A}_{q\bar q}^
{\prime}(1 \leftrightarrow 2) \right]~,
\end{equation}
where $t^a$ are the color group generators in the fundamental representation.
The amplitude ${\cal A}_{q\bar q}$ will be defined below.

Using the symmetry $(1 \leftrightarrow 2)$ of the phase space volume
(\ref{44}) and the expressions (\ref{45}), we get for the $q\bar q$ contribution
(\ref{41}) to the impact factors
$$
\Phi_{AA^{\prime}}^{cc^{\prime}(1)\{q\bar q\}}(\vec q_1, \vec q) = g^4\sum_f
\int_0^1\int\frac{d\beta}{\beta(1 - \beta)}\frac{d^{D-2}k_1}{2(2\pi)^{D-1}}
$$
\begin{equation}\label{46}
\times\left[ {C_1}_{AA^{\prime}}^{cc^{\prime}}\sum_{\lambda_1, \lambda_2}
{\cal A}_{q\bar q}{\cal A}_{q\bar q}^{\prime *} - {C_2}_{AA^{\prime}}^{cc^
{\prime}}\sum_{\lambda_1, \lambda_2}{\cal A}_{q\bar q}{\cal A}_{q\bar q}^
{\prime *}(1 \leftrightarrow 2) \right]~,
\end{equation}
where $\beta \equiv \beta_1$ and
\begin{equation}\label{47}
{C_1}_{AA^{\prime}}^{cc^{\prime}} =
tr\left(t^At^ct^{c^{\prime}}t^{A^{\prime}} +
t^At^{A^{\prime}}t^{c^{\prime}}t^c\right)~,\ \ \ \
{C_2}_{AA^{\prime}}^{cc^{\prime}} =
tr\left( t^At^ct^{A^{\prime}}
t^{c^{\prime}} +
t^At^{c^{\prime}}t^{A^{\prime}}t^c \right)~.
\end{equation}
The amplitudes ${\cal A}_{q\bar q}$ and ${\cal A}_{q\bar q}^{\prime}$ have the
forms
$$
{\cal A}_{q\bar q} = \bar u_1\frac{p\!\!\!/_B}{s}\biggl( e\!\!\!/_{A\perp}
S\!\!\!/_{12\perp} - 2(1 - \beta)\left( e_{A\perp}S_{12\perp} \right) -
e\!\!\!/_{A\perp}m_fS_{12} \biggr)v_2~,
$$
\begin{equation}\label{48}
{\cal A}_{q\bar q}^{\prime} = \bar u_1\frac{p\!\!\!/_B}{s}\biggl( e\!\!\!/_
{A^{\prime}\perp}S\!\!\!/_{12\perp}^{\prime} - 2(1 - \beta)\left( e_{A^{\prime}
\perp}S_{12\perp}^{\prime} \right) - e\!\!\!/_{A^{\prime}\perp}m_fS_{12}^
{\prime} \biggr)v_2~,
\end{equation}
with
$$
S_{12\perp} = \left( \frac{k_1}{k_{1\perp}^2 - m_f^2} - \frac{(k_1 + \beta
q_1)}{(k_1 + \beta q_1)_{\perp}^2 - m_f^2} \right)_{\perp}~,
$$
$$
S_{12\perp}^{\prime} = \left( \frac{(k_1 + \beta q)}{(k_1 + \beta q)_{\perp}^2
- m_f^2} - \frac{(k_1 + \beta q_1)}{(k_1 + \beta q_1)_{\perp}^2 - m_f^2}
\right)_{\perp}~,
$$
$$
S_{12} = \frac{1}{k_{1\perp}^2 - m_f^2} - \frac{1}{(k_1 + \beta q_1)_{\perp}^2
- m_f^2}~,
$$
\begin{equation}\label{49}
S_{12}^{\prime} = \frac{1}{(k_1 + \beta q)_{\perp}^2 - m_f^2} -
\frac{1}{(k_1 + \beta q_1)_{\perp}^2 - m_f^2}~.
\end{equation}
In these relations $u_1$ and $v_2$ are the spin wave functions of the final
quark and antiquark, respectively. The other amplitude which enters the
expression (\ref{46}) for the $q\bar q$ contribution to the impact factors is
\begin{equation}\label{410}
{\cal A}_{q\bar q}^{\prime}(1 \leftrightarrow 2) = - \bar u_1\frac{p\!\!\!/_B}
{s}\biggl( e\!\!\!/_{A^{\prime}\perp}S\!\!\!/_{21\perp}^{\prime} - 2(1 - \beta)
\left( e_{A^{\prime}\perp}S_{21\perp}^{\prime} \right) + e\!\!\!/_{A^{\prime}
\perp}m_fS_{21}^{\prime} \biggr)v_2~,
\end{equation}
with
$$
S_{21\perp}^{\prime} = S_{12\perp}^{\prime}(k_1 \leftrightarrow k_2) = \left(
- \frac{(k_1 + q_1^{\prime} + \beta q)}{(k_1 + q_1^{\prime} + \beta q)_{\perp}
^2 - m_f^2} + \frac{(k_1 + \beta q_1)}{(k_1 + \beta q_1)_{\perp}^2 - m_f^2}
\right)_{\perp}~,
$$
\begin{equation}\label{411}
S_{21}^{\prime} = S_{12}^{\prime}(k_1 \leftrightarrow k_2) = \frac{1}
{(k_1 + q_1^{\prime} + \beta q)_{\perp}^2 - m_f^2} - \frac{1}{(k_1 +
\beta q_1)_{\perp}^2 - m_f^2}~.
\end{equation}
The amplitude ${\cal A}_{q\bar q}^{\prime}(1 \leftrightarrow 2)$ is
obtained from  the amplitude ${\cal A}_{q\bar q}^{\prime}$ by 
the replacement $(1 \leftrightarrow 2)$, as it can be seen using the 
charge conjugation. An analogous expression can be written 
also for the amplitude ${\cal A}_{q\bar q}(1 \leftrightarrow 2)$, but we do not
give it here, because it does not enter Eq.~(\ref{46}). Then convolutions
in Eq.~(\ref{46}) are calculated immediately and give
$$
\sum_{\lambda_1, \lambda_2}{\cal A}_{q\bar q}{\cal A}_{q\bar q}^{\prime *} =
2\beta(1 - \beta)e_{A^{\prime}\mu}^{*\perp}e_{A\nu}^{\perp}\biggl[ g_{\perp
\perp}^{\mu\nu}\biggl( \left( S_{12\perp}^{\prime}S_{12\perp} \right)
$$
$$
- m_f^2S_{12}^{\prime}S_{12} \biggr) - S_{12\perp}^{\prime\:\nu}S_{12\perp}^{\mu}
+ \left( 1 - 4\beta(1 - \beta) \right)S_{12\perp}^{\prime\:\mu}S_{12\perp}^{\nu}
\biggr]~,
$$
$$
- \sum_{\lambda_1, \lambda_2}{\cal A}_{q\bar q}{\cal A}_{q\bar q}^{\prime *}
(1 \leftrightarrow 2) = 2\beta(1 - \beta)e_{A^{\prime}\mu}^{*\perp}e_{A\nu}^
{\perp}\biggl[ g_{\perp\perp}^{\mu\nu}\biggl( \left( S_{21\perp}^{\prime}S_
{12\perp} \right)
$$
\begin{equation}\label{412}
+ m_f^2S_{21}^{\prime}S_{12} \biggl) - S_{21\perp}^{\prime\:\nu}S_{12\perp}^{\mu}
+ \left( 1 - 4\beta(1 - \beta) \right)S_{21\perp}^{\prime\:\mu}S_{12\perp}^{\nu}
\biggr]~.
\end{equation}
This result allows us to present the quark-antiquark contribution
(\ref{41}) 
to the impact factors in the following form:
\begin{equation}\label{413}
\Phi_{AA^{\prime}}^{cc^{\prime}(1)\{q\bar q\}}(\vec q_1, \vec q) =
{C_1}_{AA^{\prime}}^{cc^{\prime}}I_1 + {C_2}_{AA^{\prime}}^{cc^{\prime}}I_2~,
\end{equation}
with
$$
I_1 = g^4e_{A^{\prime}\mu}^{*\perp}e_{A\nu}^{\perp}\sum_f
\biggl[ g_{\perp\perp}^{\mu\nu}\biggl( \int_0^1d\beta\int\frac{d^{D-2}k_1}
{(2\pi)^{D-1}}\left(S_{12\perp}^{\prime}S_{12\perp} \right) - m_f^2
\int_0^1d\beta\int\frac{d^{D-2}k_1}{(2\pi)^{D-1}}S_{12}^{\prime}S_{12} \biggl)
$$
\begin{equation}\label{414}
- \int_0^1d\beta\int\frac{d^{D-2}k_1}{(2\pi)^{D-1}}S_{12\perp}^{\prime\:\nu}
S_{12\perp}^{\mu} + \int_0^1d\beta\left( 1 - 4\beta(1 - \beta) \right)\int\frac
{d^{D-2}k_1}{(2\pi)^{D-1}}S_{12\perp}^{\prime\:\mu}S_{12\perp}^{\nu} \biggr]
\end{equation}
and
$$
I_2 = g^4e_{A^{\prime}\mu}^{*\perp}e_{A\nu}^{\perp}\sum_f
\biggl[ g_{\perp\perp}^{\mu\nu}\biggl( \int_0^1d\beta\int
\frac{d^{D-2}k_1}{(2\pi)^{D-1}}\left(S_{21\perp}^{\prime}S_{12\perp}
\right) +m_f^2\int_0^1d\beta\int\frac{d^{D-2}k_1}{(2\pi)^{D-1}}S_{21}^{\prime}
S_{12} \biggl)
$$
\begin{equation}\label{415}
- \int_0^1d\beta\int\frac{d^{D-2}k_1}{(2\pi)^{D-1}}S_{21\perp}^{\prime\:\nu}
S_{12\perp}^{\mu} + \int_0^1d\beta\left( 1 - 4\beta(1 - \beta) \right)\int\frac
{d^{D-2}k_1}{(2\pi)^{D-1}}S_{21\perp}^{\prime\:\mu}S_{12\perp}^{\nu} \biggr]~.
\end{equation}
Let us consider first $I_1$. Using the relations
$$
\int\frac{d^{D-2}k_1}{(2\pi)^{D-1}}S_{12\perp}^{\prime\:\nu}S_{12\perp}^{\mu} =
\int\frac{d^{D-2}k_1}{(2\pi)^{D-1}}S_{12\perp}^{\prime\:\mu}S_{12\perp}^{\nu}
= \beta^{2\epsilon}\biggl[ J_1^{\mu\nu}\left( q_{\perp},
\frac{m_f^2}{\beta^2} \right)
$$
\begin{equation}\label{416}
+ J_1^{\mu\nu}\left( 0_{\perp},\frac{m_f^2}{\beta^2} \right) - J_1^{\mu\nu}
\left( q_{1\perp}^{\prime},\frac{m_f^2}{\beta^2} \right) - J_1^{\mu\nu}\left(
q_{1\perp},\frac{m_f^2}{\beta^2} \right) \biggr]~,
\end{equation}
$$
J_1^{\mu\nu}\left( v_{\perp},\frac{m_f^2}{\beta^2} \right) = 
\int\frac{d^{D-2}k}{(2\pi)^{D-1}}\frac{k_{\perp}^{\mu}
(k - v)_{\perp}^{\nu}}{\left( k_{\perp}^2-\frac{m_f^2}{\beta^2} \right)
\left( (k - v)_{\perp}^2 -\frac{m_f^2}{\beta^2} \right)} =
$$
$$
\beta^{-2\epsilon}\frac{\Gamma(1 - \epsilon)}{(1 + \epsilon)
(4\pi)^{2 + \epsilon}}\biggl[ g_{\perp\perp}^{\mu\nu}\int_0^1dx
\left( m_f^2 + \beta^2x(1 - x)\vec v^{\:2} \right)^{\epsilon}
\left( \frac{1 + 2\epsilon}{\epsilon} - \frac{m_f^2}{(m_f^2 +
\beta^2x(1 - x)\vec v^{\:2})} \right)
$$
\begin{equation}\label{417}
- T_{\perp\perp}^{\mu\nu}(v_{\perp})\int_0^1dx\left( m_f^2 + \beta^2x
(1 - x)\vec v^{\:2} \right)^{\epsilon}\left( 1 - \frac{m_f^2}{(m_f^2 +
\beta^2x(1 - x)\vec v^{\:2})} \right) \biggr]~,
\end{equation}
$$
\int\frac{d^{D-2}k_1}{(2\pi)^{D-1}}S_{12}^{\prime}S_{12} =
\beta^{2\epsilon - 2}\biggl[ J_1\left( \vec q^{\:2},\frac{m_f^2}{\beta^2} \right)
+ J_1\left( 0,\frac{m_f^2}{\beta^2} \right)
$$
\begin{equation}\label{418}
- J_1\left( \vec q_1^{\:\prime\:2},\frac{m_f^2}{\beta^2} \right) - J_1\left(
\vec q_1^{\:2},\frac{m_f^2}{\beta^2} \right) \biggr]~,
\end{equation}
$$
J_1\left( \vec v^{\:2},\frac{m_f^2}{\beta^2} \right) = \int\frac{d^{D-2}k}
{(2\pi)^{D-1}}\frac{1}{\left( \vec k^{\:2} +\frac{m_f^2}{\beta^2} \right)\left(
(\vec k - \vec v)^2 +\frac{m_f^2}{\beta^2} \right)} =
$$
\begin{equation}\label{419}
2\beta^{2 - 2\epsilon}\frac{\Gamma(1 - \epsilon)}{(4\pi)^{2 +
\epsilon}}\int_0^1dx\left( m_f^2 + \beta^2x(1 - x)\vec v^{\:2}
\right)^{\epsilon}\frac{1}{(m_f^2 + \beta^2x(1 - x)\vec v^{\:2})}~,
\end{equation}
together with Eqs.~(\ref{35}), (\ref{39}) and (\ref{414}), performing also 
the following change of the integration variables in Eqs.~(\ref{414}) and
(\ref{416})-(\ref{419}):
\begin{equation}\label{420}
x_1 = \beta x~,\ \ \ \ \ \ x_2 = \beta (1 - x)~,
\end{equation}
we can set $I_1$ in the form
$$
I_1 = -g^4\frac{\Gamma(1 - \epsilon)}{(1 + \epsilon)(4\pi)^
{2 + \epsilon}}e_{A^{\prime}\perp}^{*\mu}e_{A\perp}^{\nu}\biggr[ g_{\mu\nu}^
{\perp\perp}\biggl( F_1^{(+)}\left( \vec q_1^{\:\prime\:2} \right) + F_1^{(+)}
\left( \vec q_1^{\:2} \right) - F_1^{(+)}\left( \vec q^{\:2} \right) \biggr)
$$
\begin{equation}\label{421}
+ 4T_{\mu\nu}^{\perp\perp}\left( q_{1\perp}^{\prime} \right)F_1^{(-)}\left(
\vec q_1^{\:\prime\:2} \right) + 4T_{\mu\nu}^{\perp\perp}\left( q_{1\perp}
\right)F_1^{(-)}\left( \vec q_1^{\:2} \right) - 4T_{\mu\nu}^{\perp\perp}\left(
q_{\perp} \right)F_1^{(-)}\left( \vec q^{\:2} \right) \biggr]~.
\end{equation}

To calculate $I_2$ we use the following relations which can be obtained
starting from Eqs.~(\ref{49}) and (\ref{411}):
$$
\int\frac{d^{D-2}k_1}{(2\pi)^{D-1}}S_{21\perp}^{\prime\:\mu}S_{12\perp}^{\nu}
= \int\frac{d^{D-2}k_1}{(2\pi)^{D-1}}S_{21\perp}^{\prime\:\nu}S_{12\perp}^{\mu}
= \frac{\Gamma(1 - \epsilon)}{(4\pi)^{2 + \epsilon}}\biggl\{
\biggr[ g_{\perp\perp}^{\mu\nu}\biggl( - \frac{1}{\epsilon}\left(m_f^2 
\right)^{\epsilon}
$$
$$
+ \frac{1}{\epsilon}\int_0^1dx\left( m_f^2 + x(1 - x)\vec v^{\:2}
\right)^{\epsilon} \biggr) + 2\int_0^1dx\left( m_f^2 + x(1 - x)\vec
v^{\:2} \right)^{\epsilon}
$$
\begin{equation}\label{422}
\times\frac{x(1 - x)\vec v^{\:2}}{\left( m_f^2 + x(1 - x)\vec v^{\:2} \right)}
\frac{v_{\perp}^{\mu}v_{\perp}^{\nu}}{v_{\perp}^2} \biggr]\biggl|_{v_{\perp} =
(1 - \beta)q_{1\perp}^{\prime}} + \biggl[ \cdots \biggr]\biggl|_{v_{\perp} =
\beta q_{1\perp}} - \biggl[ \cdots \biggr]\biggl|_{v_{\perp} = \left( (1 -
\beta)q_1^{\prime} + \beta q_1 \right)_{\perp}} \biggr\}~,
\end{equation}
$$
\int\frac{d^{D-2}k_1}{(2\pi)^{D-1}}m_f^2S_{21}^{\prime}S_{12} = 
\frac{2\Gamma(1 - \epsilon)}{(4\pi)^{2 + \epsilon}}\biggl\{ \biggl[
\left( m_f^2 \right)^{\epsilon} - \int_0^1dx\left( m_f^2
+ x(1 - x)\vec v^{\:2} \right)^{\epsilon}
$$
\begin{equation}\label{423}
\times\frac{m_f^2}{\left( m_f^2 + x(1 - x)\vec v^{\:2} \right)} \biggr]\biggl|
_{\vec v = (1 - \beta)\vec q_1^{\:\prime}} + \biggl[ \cdots \biggr]\biggl|
_{\vec v = \beta\vec q_1} - \biggl[ \cdots \biggr]\biggl|_{\vec v = (1 - \beta)
\vec q_1^{\:\prime} + \beta\vec q_1} \biggr\}~.
\end{equation}
Then we get from Eq.~(\ref{415})
$$
I_2 = 2g^4\frac{\Gamma(1 - \epsilon)}{(4\pi)^{2 + \epsilon}}
e_{A^{\prime}\perp}^{*\mu}e_{A\perp}^{\nu}\sum_f\biggl\{ \biggl[ g_{\mu\nu}^
{\perp\perp}\biggl( - \frac{2}{3\epsilon}\left( m_f^2 \right)
^{\epsilon} + \int_0^1\int_0^1d\beta dx \left( m_f^2 + x(1 - x)\vec v
^{\:2} \right)^{\epsilon}
$$
$$
\times\left( \frac{1 + 2\epsilon - 2\beta(1 - \beta)}{\epsilon} - \frac{2m_f^2}
{\left( m_f^2 + x(1 - x)\vec v^{\:2} \right)} \right) \biggr)
$$
$$
- 4\int_0^1\int_0^1d\beta dx\beta(1 - \beta)\left( m_f^2 + x(1 - x)\vec
v^{\:2} \right)^{\epsilon}\frac{x(1 - x)\vec v^{\:2}}{\left( m_f^2 +
x(1 - x)\vec v^{\:2} \right)}
$$
\begin{equation}\label{424}
\times\frac{v_{\perp\mu}v_{\perp\nu}}{v_{\perp}^2} \biggr]\biggl|_{v_{\perp} =
(1 - \beta)q_{1\perp}^{\prime}} + \biggl[ \cdots \biggr]\biggl|_{v_{\perp} =
\beta q_{1\perp}} - \biggl[ \cdots \biggr]\biggl|_{v_{\perp} = \left( (1 -
\beta)q_1^{\prime} + \beta q_1 \right)_{\perp}} \biggr\}~.
\end{equation}
Finally, the quark-antiquark contribution to the gluon impact factors is
given by 
Eq.~(\ref{413}) with the color factors ${C_{1,2}}_{AA^{\prime}}^{cc^{\prime}}$
defined in Eq.~(\ref{47}), $I_1$ and $I_2$ expressed in Eqs.~(\ref{421}) 
and (\ref{424}), respectively, the functions $F_1^{(\pm)}$ being the same as
in the previous section (see Eq.~(\ref{39})). 

\section{Two-gluon contribution}
\setcounter{equation}{0}
The two-gluon contribution to the gluon impact factors defined in Eq.~(\ref{28}) 
can be presented as 
\begin{equation}\label{51}
\Phi_{AA^{\prime}}^{cc^{\prime}(1)\{GG\}}(\vec q_1, \vec q; s_0) = \sum_
{\stackrel{\lambda_1, \lambda_2}{i_1, i_2}}\int\theta(s_{\Lambda} - s_{AR})
\frac{ds_{AR}\:d\rho_{\{GG\}}}{(2\pi)}\Gamma^c_{\{GG\}A}\left( \Gamma^
{c^{\prime}}_{\{GG\}A^{\prime}} \right)^*~,
\end{equation}
where $\lambda_1$, $\lambda_2$ and $i_1$, $i_2$ are helicities and color
indices of the produced gluons with momenta $k_1$ and $k_2$. The expressions for
the Sudakov representation of the produced gluon momenta and their
invariant mass are the same as in the quark-antiquark case (see Eqs.~(\ref{42})
and (\ref{43})), but with $m_f = 0$. As
for the integration volume element, we should introduce in Eq.~(\ref{44})
the factors $\theta(s_{\Lambda} - s_{AR})$ and $1/2!$ due to the gluons 
identity. The two-gluon production effective amplitude has the form~\cite{3}
\begin{equation}\label{52}
\Gamma^c_{\{GG\}A} = 4g^2q_1^2\left[ T_{i_1A}^{c_1}T_{i_2c}^{c_1}
{\cal A}_{GG} + (1 \leftrightarrow 2) \right]~,
\end{equation}
while the other amplitude 
$\Gamma^{c^{\prime}}_{\{GG\}A^{\prime}}$
in Eq.~(\ref{51}) can be obtained from this relation by the evident 
substitutions
$A \rightarrow A^{\prime},\ c \rightarrow c^{\prime}$. The gauge 
invariant expression for the amplitude
${\cal A}_{GG}$ is  rather complicated~\cite{3}, but in our gauge 
(\ref{215}) it becomes very simple~\cite{14} (see below).  Just as in the
quark-antiquark case, we can use the $(1 \leftrightarrow 2)$ symmetry of the
integration volume element to obtain
$$
\Phi_{AA^{\prime}}^{cc^{\prime}(1)\{GG\}}(\vec q_1, \vec q; s_0) = 8g^4
q_1^2q_1^{\prime\:2}\int_0^1\int\theta(s_{\Lambda} - s_{AR})\frac{d\beta}
{\beta(1 - \beta)}\frac{d^{D-2}k_1}{(2\pi)^{D-1}}
$$
\begin{equation}\label{53}
\times\left[ {C_3}_{AA^{\prime}}^{cc^{\prime}}\sum_{\lambda_1, \lambda_2}
{\cal A}_{GG}{\cal A}_{GG}^{\prime *} + {C_4}_{AA^{\prime}}^{cc^{\prime}}
\sum_{\lambda_1, \lambda_2}{\cal A}_{GG}{\cal A}_{GG}^{\prime *}
(1 \leftrightarrow 2) \right]~,
\end{equation}
where $\beta \equiv \beta_1$ and
\begin{equation}\label{54}
{C_3}_{AA^{\prime}}^{cc^{\prime}} = \left( T^{c_1^
{\prime}}T^{c_1} \right)_{A^{\prime}A}\left( T^{c_1^{\prime}}T^{c_1} \right)
_{c^{\prime}c},\ \ \ \ \ {C_4}_{AA^{\prime}}^{cc^{\prime}} =
\left( T^{c_1^{\prime}}T^{c_1} \right)_{c^{\prime}A}\left( T^{c_1^{\prime}}
T^{c_1} \right)_{A^{\prime}c}~.
\end{equation}
The amplitudes ${\cal A}_{GG}$, ${\cal A}_{GG}^{\prime}$ and
${\cal A}_{GG}^{\prime}(1 \leftrightarrow 2)$ in the gauge
(\ref{215}, \ref{216}) have the forms
$$
{\cal A}_{GG} = \frac{1}{2q_1^2}\biggl[ - \beta(1 - \beta)\left( e_{1\perp}
^*e_{2\perp}^* \right)\left( e_{A\perp}{\widetilde R}_{12\perp} \right)
$$
$$
+ \beta\left( e_{1\perp}^*e_{A\perp} \right)\left( e_{2\perp}^*{\widetilde R}
_{12\perp} \right) + (1 - \beta)\left( e_{2\perp}^*e_{A\perp} \right)\left(
e_{1\perp}^*{\widetilde R}_{12\perp} \right) \biggr]~,
$$
$$
{\cal A}_{GG}^{\prime} = \frac{1}{2q_1^{\prime\:2}}\biggl[ - \beta(1 - \beta)
\left( e_{1\perp}^*e_{2\perp}^* \right)\left( e_{A^{\prime}\perp}{\widetilde R}
_{12\perp}^{\prime} \right)
$$
$$
+ \beta\left( e_{1\perp}^*e_{A^{\prime}\perp} \right)\left( e_{2\perp}^*
{\widetilde R}_{12\perp}^{\prime} \right) + (1 - \beta)\left( e_{2\perp}^*
e_{A^{\prime}\perp} \right)\left( e_{1\perp}^*{\widetilde R}_{12\perp}
^{\prime} \right) \biggr]~,
$$
$$
{\cal A}_{GG}^{\prime}(1 \leftrightarrow 2) = \frac{1}{2q_1^{\prime\:2}}\biggl[
- \beta(1 - \beta)\left( e_{1\perp}^*e_{2\perp}^* \right)\left( e_{A^{\prime}
\perp}{\widetilde R}_{21\perp}^{\prime} \right)
$$
\begin{equation}\label{55}
+ \beta\left( e_{1\perp}^*e_{A^{\prime}\perp} \right)\left( e_{2\perp}^*
{\widetilde R}_{21\perp}^{\prime} \right) + (1 - \beta)\left( e_{2\perp}^*
e_{A^{\prime}\perp} \right)\left( e_{1\perp}^*{\widetilde R}_{21\perp}
^{\prime} \right) \biggr]~,
\end{equation}
where $e_1$ and $e_2$ are the polarization vectors of the produced gluons and
$$
{\widetilde R}_{12\perp} = \left( \frac{k_1}{k_{1\perp}^2} - \frac{(k_1 + \beta
q_1)}{(k_1 + \beta q_1)_{\perp}^2} \right)_{\perp}~,\ \ \ \ \ \ \ 
{\widetilde R}_{12\perp}^{\prime} = \left( \frac{(k_1 + \beta q)}{(k_1 + 
\beta q)_{\perp}^2} - \frac{(k_1 + \beta q_1)}{(k_1 + \beta q_1)_{\perp}^2}
\right)_{\perp}~,
$$
\begin{equation}\label{56}
{\widetilde R}_{21\perp}^{\prime} = {\widetilde R}_{12\perp}^{\prime}
(1 \leftrightarrow 2) = \left( - \frac{(k_1 + q_1^{\prime} + \beta q)}{(k_1
+q_1^{\prime} + \beta q)_{\perp}^2} + \frac{(k_1 + \beta q_1)}{(k_1 + \beta
q_1)_{\perp}^2} \right)_{\perp}.
\end{equation}
Eqs.~(\ref{55}) leads to the following expressions for the convolutions in the
relation (\ref{53}) for the two-gluon contribution to the impact factors:
$$
\sum_{\lambda_1, \lambda_2}{\cal A}_{GG}{\cal A}_{GG}^{\prime *} = \frac{1}
{4q_1^2q_1^{\prime\:2}}\biggl[ \left( \beta^2 + (1 - \beta)^2 \right)\left(
e_{A^{\prime}\perp}^*e_{A\perp} \right)g_{\mu\nu}^{\perp\perp}
$$
$$
+ 2\beta(1 - \beta)e_{A\mu}^{\perp}e_{A^{\prime}\nu}^{*\perp} - 2\beta
(1 - \beta)\left( 1 - (1 + \epsilon)\beta(1 - \beta) \right)e_{A^{\prime}\mu}
^{*\perp}e_{A\nu}^{\perp} \biggr]{\widetilde R}_{12\perp}^{\prime\:\mu}
{\widetilde R}_{12\perp}^{\nu}~,
$$
$$
\sum_{\lambda_1, \lambda_2}{\cal A}_{GG}{\cal A}_{GG}^{\prime *}
(1 \leftrightarrow 2) = \frac{1}{4q_1^2q_1^{\prime\:2}}\biggl[ \left( \beta^2
+ (1 - \beta)^2 \right)\left( e_{A^{\prime}\perp}^*e_{A\perp} \right)
g_{\mu\nu}^{\perp\perp}
$$
\begin{equation}\label{57}
+ 2\beta(1 - \beta)e_{A\mu}^{\perp}e_{A^{\prime}\nu}^{*\perp} - 2\beta
(1 - \beta)\left( 1 - (1 + \epsilon)\beta(1 - \beta) \right)e_{A^{\prime}\mu}
^{*\perp}e_{A\nu}^{\perp} \biggr]{\widetilde R}_{21\perp}^{\prime\:\mu}
{\widetilde R}_{12\perp}^{\nu}~,
\end{equation}
which allow us to get through Eqs.~(\ref{53}) the relation
$$
\Phi_{AA^{\prime}}^{cc^{\prime}(1)\{GG\}}(\vec q_1, \vec q; s_0) = 2g^4
\int_0^1\int\theta(s_{\Lambda} - s_{AR})\frac{d\beta}{\beta(1 - \beta)}
\frac{d^{D-2}k_1}{(2\pi)^{D-1}}\biggl[ \left( \beta^2 + (1 - \beta)^2 \right)
$$
$$
\times\left( e_{A^{\prime}\perp}^*e_{A\perp} \right)g_{\mu\nu}^{\perp\perp}
+ 2\beta(1 - \beta)e_{A\mu}^{\perp}e_{A^{\prime}\nu}^{*\perp} - 2\beta(1 -
\beta)\left( 1 - (1 + \epsilon)\beta(1 - \beta) \right)e_{A^{\prime}\mu}^
{*\perp}e_{A\nu}^{\perp} \biggr]
$$
\begin{equation}\label{58}
\times\biggl[ {C_3}_{AA^{\prime}}^{cc^{\prime}}{\widetilde R}_{12\perp}^
{\prime\:\mu}{\widetilde R}_{12\perp}^{\nu} + {C_4}_{AA^{\prime}}^{cc^{\prime}}
{\widetilde R}_{21\perp}^{\prime\:\mu}{\widetilde R}_{12\perp}^{\nu} \biggr]~.
\end{equation}
It seems to be convenient now to change the integration momentum $k_1$ in
the following way:
\begin{equation}\label{59}
k_1 \rightarrow - \beta k_1~,\ \ \ \ \ \
d^{D-2}k_1 \rightarrow \beta^{2 + 2\epsilon}d^{D-2}k_1~,
\end{equation}
which states the substitutions
$$
s_{AR} \rightarrow \frac{\beta(\vec k_1 - \vec q_1)^2}{(1 - \beta)}~,\ \ \ \ \ 
{\widetilde R}_{12\perp} \rightarrow - \frac{1}{\beta}R_{12\perp} = - \frac
{1}{\beta}\left( \frac{k_1}{k_{1\perp}^2} - \frac{(k_1 - q_1)}{(k_1 - q_1)
_{\perp}^2} \right)_{\perp}~,
$$
$$
{\widetilde R}_{12\perp}^{\prime} \rightarrow - \frac{1}{\beta}R_{12\perp}
^{\prime} = - \frac{1}{\beta}\left( \frac{(k_1 - q)}{(k_1 - q)_{\perp}^2} -
\frac{(k_1 - q_1)}{(k_1 - q_1)_{\perp}^2} \right)_{\perp}~,
$$
\begin{equation}\label{510}
{\widetilde R}_{21\perp}^{\prime} \rightarrow - \frac{1}{\beta}R_{21\perp}
^{\prime} = - \frac{1}{\beta}\left( - \frac{\beta(\beta k_1 - q_1^{\prime}
- \beta q)}{(\beta k_1 - q_1^{\prime} - \beta q)_{\perp}^2} + \frac
{(k_1 - q_1)}{(k_1 - q_1)_{\perp}^2} \right)_{\perp}~.
\end{equation}
Consequently, the representation for the two-gluon part of the impact
factors reads
$$
\Phi_{AA^{\prime}}^{cc^{\prime}(1)\{GG\}}(\vec q_1, \vec q; s_0) = 2g^4
\int_0^1\int\theta\left( 1 - \frac{\beta(\vec k_1 - \vec q_1)^2}
{(1 - \beta)s_{\Lambda}} \right)\frac{\beta^{2\epsilon -1}d\beta}{(1 - \beta)}
\frac{d^{D-2}k_1}{(2\pi)^{D-1}}
$$
$$
\times\biggl[ \left( \beta^2 + (1 - \beta)^2 \right)\left( e_{A^{\prime}\perp}
^*e_{A\perp} \right)g_{\mu\nu}^{\perp\perp} + 2\beta(1 - \beta)e_{A\mu}^{\perp}
e_{A^{\prime}\nu}^{*\perp} - 2\beta(1 - \beta)
$$
\begin{equation}\label{511}
\times\left( 1 - (1 + \epsilon)\beta(1 - \beta) \right)e_{A^{\prime}\mu}^
{*\perp}e_{A\nu}^{\perp} \biggr]\biggl[ {C_3}_{AA^{\prime}}^{cc^{\prime}}
R_{12\perp}^{\prime\:\mu}R_{12\perp}^{\nu} + {C_4}_{AA^{\prime}}^{cc^{\prime}}
R_{21\perp}^{\prime\:\mu}R_{12\perp}^{\nu} \biggr]~.
\end{equation}
The lower limit of integration over $\beta$ is not affected by the 
$\theta$-function in the integrand of the R.H.S. of this equation and can
be put equal to zero. The integral over $\beta$ is convergent in this lower
limit when the parameter
$s_{\Lambda}$ goes to infinity (there is, however, the usual infrared 
divergence at 
 $\beta = 0$, $\epsilon = 0$, but it is irrelevant for the separation of
the QMRK and MRK contributions determined by the parameter $s_{\Lambda}$).
As for the upper limit of the integration, we easily get 
\begin{equation}\label{512}
\beta_{max} = 1 - \frac{(\vec k_1 - \vec q_1)^2}{s_{\Lambda}}~.
\end{equation}
There are two factors ${\widetilde I}_3$ and $I_4$ in front of the two color
structures ${C_3}_{AA^{\prime}}^{cc^{\prime}}$ and
${C_4}_{AA^{\prime}}^{cc^{\prime}}$, respectively, in Eq.~(\ref{511}). 
The first is
$$
{\widetilde I}_3 = 2g^4\int\int_0^{\beta_{max}}\frac{\beta^{2\epsilon -1}
d\beta}{(1 - \beta)}\frac{d^{D-2}k_1}{(2\pi)^{D-1}}\biggl[ \left( \beta^2 +
(1 - \beta)^2 \right)\left( e_{A^{\prime}\perp}^*e_{A\perp} \right)g_{\mu\nu}
^{\perp\perp}
$$
\begin{equation}\label{513}
+ 2\beta(1 - \beta)e_{A\mu}^{\perp}e_{A^{\prime}\nu}^{*\perp} - 2\beta(1 -
\beta)\left( 1 - (1 + \epsilon)\beta(1 - \beta) \right)e_{A^{\prime}\mu}^
{*\perp}e_{A\nu}^{\perp} \biggr]R_{12\perp}^{\prime\:\mu}R_{12\perp}^{\nu}~,
\end{equation}
while the second can be obtained from ${\widetilde I}_3$ by the replacement
\begin{equation}\label{514}
R_{12\perp}^{\prime} \rightarrow R_{21\perp}^{\prime}~,\ \ \ \ \ \ \beta_{max}
\rightarrow 1~.
\end{equation}
Notice that the second of the substitutions  (\ref{514}) is possible because 
the integration over $\beta$ in the expression for $I_4$ is convergent. 
Let us firstly consider the factor ${\widetilde I}_3$. In this case the
integration over $\beta$ can be carried out without any difficulties
(see Eqs.~(\ref{510}) and (\ref{513})) and leads to 
$$
{\widetilde I}_3 = 2g^4\int\frac{d^{D-2}k_1}{(2\pi)^{D-1}}\biggl[ \biggl(
\ln\left( \frac{s_{\Lambda}}{(\vec k_1 - \vec q_1)^2} \right) + \frac{(1 -
2\epsilon)}{2\epsilon(1 + 2\epsilon)} + \psi(1) - \psi(1 + 2\epsilon) \biggr)
$$
\begin{equation}\label{515}
\times\left( e_{A^{\prime}\perp}^*e_{A\perp} \right)g_{\mu\nu}^{\perp\perp}
+ \frac{2}{(1 + 2\epsilon)}e_{A\mu}^{\perp}e_{A^{\prime}\nu}^{*\perp} -
\frac{(5 + 2\epsilon)}{(1 + 2\epsilon)(3 + 2\epsilon)}e_{A^{\prime}\mu}
^{*\perp}e_{A\nu}^{\perp} \biggr]R_{12\perp}^{\prime\:\mu}R_{12\perp}^{\nu}~.
\end{equation}
Now we should also include the counterterm given by the second term in the 
R.H.S. of Eq.~(\ref{28}) into the two-gluon contribution
to the impact factors, as it was already mentioned in Section 2. Using
Eqs.~(\ref{214}) and (\ref{312}), it is easy to check the following
relation for the counterterm, taken with the minus sign which appears in 
Eq.~(\ref{28}):
\begin{equation}\label{516}
{\mbox{counterterm}} = {C_3}_{AA^{\prime}}^{cc^{\prime}}2g^4
\int\frac{d^{D-2}k_1}
{(2\pi)^{D-1}}\biggl[ - \frac{1}{2}\ln\left( \frac{s_{\Lambda}^2}
{s_0(\vec k_1 - \vec q_1)^2} \right)\left( e_{A^{\prime}\perp}^*e_{A\perp}
\right)g_{\mu\nu}^{\perp\perp} \biggr]R_{12\perp}^{\prime\:\mu}
R_{12\perp}^{\nu}~.
\end{equation}
Therefore, we can make the redefinition ${\widetilde I}_3 \rightarrow I_3$ in 
order to include the counterterm contribution:
$$
{\widetilde I}_3 \rightarrow I_3 = 2g^4\int\frac{d^{D-2}k_1}{(2\pi)^{D-1}}
\biggl[ \biggl( \frac{1}{2}\ln\left( \frac{s_0}{(\vec k_1 - \vec q_1)^2}
\right) + \frac{(1 - 2\epsilon)}{2\epsilon(1 + 2\epsilon)} + \psi(1) - \psi
(1 + 2\epsilon) \biggr)
$$
\begin{equation}\label{517}
\times\left( e_{A^{\prime}\perp}^*e_{A\perp} \right)g_{\mu\nu}^{\perp\perp} +
\frac{2}{(1 + 2\epsilon)}e_{A\mu}^{\perp}e_{A^{\prime}\nu}^{*\perp} - \frac
{(5 + 2\epsilon)}{(1 + 2\epsilon)(3 + 2\epsilon)}e_{A^{\prime}\mu}^{*\perp}
e_{A\nu}^{\perp} \biggr]R_{12\perp}^{\prime\:\mu}R_{12\perp}^{\nu}~.
\end{equation}
Then the full expression for the two-gluon contribution to the gluon impact
factors
(\ref{28}) does not depend on the artificial parameter $s_{\Lambda}$ and 
takes the form
\begin{equation}\label{518}
\Phi_{AA^{\prime}}^{cc^{\prime}(1)\{GG\}}(\vec q_1, \vec q; s_0) =
{C_3}_{AA^{\prime}}^{cc^{\prime}}I_3 + {C_4}_{AA^{\prime}}^{cc^{\prime}}I_4~.
\end{equation}
The next step to do is the calculation of $I_3$. With help of
Eqs.~(\ref{510}) and (\ref{517}) we obtain
$$
I_3 = - \frac{1}{2}g^4\left( e_{A^{\prime}\perp}^*e_{A\perp} \right)\biggl[
J_2(\vec q_1^{\:\prime\:2}) + J_2(\vec q_1^{\:2}) - \frac{4\Gamma(1 - \epsilon)
\Gamma^2(1 + \epsilon)}{(4\pi)^{2 + \epsilon}\Gamma(1 +
2\epsilon)}K_1 \biggr]
$$
$$
- g^4\biggl[ \biggl( \ln\left( \frac{s_0}{\vec q^{\:2}} \right) + \frac{(1 -
2\epsilon)}{\epsilon(1 + 2\epsilon)} + 2\psi(1) - 2\psi(1 +2\epsilon)\biggr)
\left( e_{A^{\prime}\perp}^*e_{A\perp} \right)g_{\mu\nu}^{\perp\perp}
$$
\begin{equation}\label{519}
+ \frac{2}{(3 + 2\epsilon)}e_{A^{\prime}\mu}^{*\perp}e_{A\nu}^{\perp} \biggr]
\biggl[ J_2^{\mu\nu}(q_{1\perp}^{\prime}) + J_2^{\mu\nu}(q_{1\perp}) -
J_2^{\mu\nu}(q_{\perp}) \biggr]~,
\end{equation}
with
$$
J_2(\vec v^{\:2}) = \int\frac{d^{D-2}k}{(2\pi)^{D-1}}\ln\left( \frac{\vec q
^{\:2}}{\vec k^{\:2}} \right)\frac{\vec v^{\:2}}{\vec k^{\:2}(\vec k - \vec v)^2}
= 2\frac{\Gamma(1 - \epsilon)}{(4\pi)^{2 + \epsilon}}\frac
{\Gamma^2(1 + \epsilon)}{\Gamma(1 + 2\epsilon)}\biggl[ \frac{2}{\epsilon}\ln
\left( \frac{\vec q^{\:2}}{\vec v^{\:2}} \right)
$$
\begin{equation}\label{520}
\times\left( \vec v^{\:2} \right)^{\epsilon} + \frac{1}{\epsilon}
\biggl( \frac{1}{\epsilon} + 2\psi(1 - \epsilon) + 2\psi(1 + 2\epsilon) -
2\psi(1) - 2\psi(1 + \epsilon) \biggr)\left( \vec v^{\:2} \right)
^{\epsilon} \biggr]~,
\end{equation}
\begin{equation}\label{521}
K_1 = \frac{(4\pi)^{2 + \epsilon}\Gamma(1 + 2\epsilon)}
{4\Gamma(1 - \epsilon)\Gamma^2(1 + \epsilon)}\int\frac{d^{D-2}k}{(2\pi)^{D-1}}
\ln\left( \frac{\vec q^{\:2}}{\vec k^{\:2}} \right)\frac{\vec q^{\:2}}{(\vec k -
\vec q_1^{\:\prime})^2(\vec k - \vec q_1)^2}~,
\end{equation}
$$
J_2^{\mu\nu}(v_{\perp}) = J_1^{\mu\nu}\left( v_{\perp},
\frac{m_f^2}{\beta^2} \right)
\biggl|_{m_f = 0} = \frac{\Gamma(1 - \epsilon)}{(4\pi)^{2 +
\epsilon}}\frac{\Gamma^2(1 + \epsilon)}{\Gamma(1 + 2\epsilon)}
$$
\begin{equation}\label{522}
\times\biggl[ \frac{1}{\epsilon(1 + \epsilon)}\left( \vec v^{\:2}
\right)^{\epsilon}g_{\perp\perp}^{\mu\nu} - \frac{1}{(1 + \epsilon)(1 +
2\epsilon)}\left( \vec v^{\:2} \right)^{\epsilon}T_{\perp\perp}
^{\mu\nu}(v_{\perp}) \biggr]~.
\end{equation}
While the integrals $J_2$ and $J_1^{\mu\nu}$ (the second given in
Eq.~(\ref{417})) have been calculated exactly, $K_1$
is  more complicated and we can calculate it only in the form of an
expansion in $\epsilon$. We do it in the Appendix.
Substituting the exact results for $J_2$ and $J_2^{\mu\nu}$, $I_3$ becomes
$$
I_3 = - 2g^4\frac{\Gamma(1 - \epsilon)}{(4\pi)^{2 + \epsilon}}
\frac{\Gamma^2(1 + \epsilon)}{\Gamma(1 + 2\epsilon)}
e_{A^{\prime}\mu}^{*\perp}
e_{A\nu}^{\perp}\biggl\{ g_{\perp\perp}^{\mu\nu}\biggl[ F_2(\vec q_1
^{\:\prime\:2}) + F_2(\vec q_1^{\:2}) 
- \frac{1}{\epsilon}\biggl( \ln\left(\frac{s_0}{\vec q^{\:2}} \right) 
+ \frac{1}{\epsilon} - \frac{4}{(1 +2\epsilon)}
$$
$$
+ \frac{1}{(1 + \epsilon)(3 + 2\epsilon)} + 2\psi(1) - 2\psi(1 + 2\epsilon)
\biggr)\left( \vec q^{\:2} \right)^{\epsilon} - K_1 \biggr]
- \frac{1}{(1 + \epsilon)(1 + 2\epsilon)(3 + 2\epsilon)}
$$
\begin{equation}\label{523}
\times \biggl[ \left( \vec q_1^{\:\prime\:2} 
\right)^{\epsilon}T_{\perp\perp}^{\mu\nu}
(q_{1\perp}^{\prime}) + \left( \vec q_1^{\:2} \right)^{\epsilon}
T_{\perp\perp}^{\mu\nu}(q_{1\perp}) - \left( \vec q^{\:2} \right)
^{\epsilon}T_{\perp\perp}^{\mu\nu}(q_{\perp}) \biggr] \biggr\}~,
\end{equation}
where
$$
F_2(\vec v^{\:2}) = \frac{1}{\epsilon}\biggl( \ln\left( \frac{s_0}{\vec
v^{\:2}}
\right) + \frac{3}{2\epsilon} - \frac{4}{(1 + 2\epsilon)} + \frac{1}{(1 + 
\epsilon)(3 + 2\epsilon)}
$$
\begin{equation}\label{524}
+ \psi(1) + \psi(1 - \epsilon) - \psi(1 + \epsilon) - \psi(1 + 2\epsilon)
\biggr)\left( \vec v^{\:2} \right)^{\epsilon}~.
\end{equation}
The calculation of $I_4$ is quite straightforward and can be performed
starting from its definition through Eqs.~(\ref{513}) and (\ref{514}). We 
obtain
$$
I_4 = 2g^4\int_0^1\frac{\beta^{2\epsilon}d\beta}{\beta(1 - \beta)}\biggl[
\left( \beta^2 + (1 - \beta)^2 \right)\left( e_{A^{\prime}\perp}^*e_{A\perp}
\right)g_{\mu\nu}^{\perp\perp} + 2\beta(1 - \beta)
e_{A\mu}^{\perp}e_{A^{\prime}\nu}^{*\perp}
$$
\begin{equation}\label{525}
- 2\beta(1 - \beta)\left( 1 - (1 + \epsilon)\beta(1 - \beta) \right)
e_{A^{\prime}\mu}^{*\perp}e_{A\nu}^{\perp} \biggr]\int\frac{d^{D-2}k_1}
{(2\pi)^{D-1}}R_{21\perp}^{\prime\:\mu}R_{12\perp}^{\nu}~.
\end{equation}
This equation, using Eqs.~(\ref{510}) and (\ref{522}), leads to
$$
I_4 = 2g^4\int_0^1\frac{d\beta}{\beta(1 - \beta)}\biggl[ \left( \beta^2 +
(1 - \beta)^2 \right)\left( e_{A^{\prime}\perp}^*e_{A\perp} \right)g_{\mu\nu}
^{\perp\perp} + 2(1 + \epsilon)\left( \beta(1 - \beta) \right)^2e_{A^{\prime}
\mu}^{*\perp}e_{A\nu}^{\perp} \biggr]
$$
\begin{equation}\label{526}
\times\biggl[ (1 - \beta)^{2\epsilon}J_2^{\mu\nu}(q_{1\perp}^{\prime}) +
\beta^{2\epsilon}J_2^{\mu\nu}(q_{1\perp}) - J_2^{\mu\nu}\left( (1 - \beta)
q_{1\perp}^{\prime} + \beta q_{1\perp} \right) \biggr]~,
\end{equation}
which can be put in the form
$$
I_4 = - 2g^4\frac{\Gamma(1 - \epsilon)}{(4\pi)^{2 + \epsilon}}
\frac{\Gamma^2(1 + \epsilon)}{\Gamma(1 + 2\epsilon)}
e_{A^{\prime}\mu}^{*\perp}
e_{A\nu}^{\perp}\biggl[ g_{\perp\perp}^{\mu\nu}\frac{(11 + 8\epsilon)}
{\epsilon(1 + 2\epsilon)(3 + 2\epsilon)}\biggl( \left( \vec q_1
^{\:\prime\:2} \right)^{\epsilon} + \left( \vec q_1^{\:2}
\right)^{\epsilon} \biggr)
$$
$$
- \frac{2}{(1 + 2\epsilon)(3 + 2\epsilon)}\biggl( \left( \vec q_1
^{\:\prime\:2} \right)^{\epsilon}\frac{q_{1\perp}^{\prime\:\mu}q_{1\perp}
^{\prime\:\nu}}{q_{1\perp}^{\prime\:2}} + \left( \vec q_1^{\:2}
\right)^{\epsilon}\frac{q_{1\perp}^{\mu}q_{1\perp}^{\nu}}{q_{1\perp}^2} \biggr)
$$
\begin{equation}\label{527}
+ g_{\perp\perp}^{\mu\nu}\biggl( \frac{2}{\epsilon}K_2 - \frac{4}{\epsilon}K_3
+ \frac{2(1 + \epsilon)}{\epsilon(1 + 2\epsilon)}K_4 \biggr) + \frac{4(1 +
\epsilon)}{(1 + 2\epsilon)}K_{4\perp\perp}^{\mu\nu} \biggr]~.
\end{equation}
The integrals
$$
K_2 = \int_0^1\frac{d\beta}{\beta(1 - \beta)}\biggl( \left( \left(
(1 - \beta)\vec q_1^{\:\prime} + \beta\vec q_1 \right)^2 \right)
^{\epsilon} - (1 - \beta)^{2\epsilon}\left( \vec q_1^{\:\prime\:2} 
\right)^{\epsilon} - \beta^{2\epsilon}\left( \vec q_1^{\:2} 
\right)^{\epsilon} \biggr)~,
$$
$$
K_3 = \int_0^1d\beta\left( \left( (1 - \beta)\vec q_1^{\:\prime} + \beta
\vec q_1 \right)^2 \right)^{\epsilon}~,\ \ \ \ \ \
K_4 = \int_0^1d\beta\beta
(1 - \beta)\left( \left( (1 - \beta)\vec q_1^{\:\prime} + \beta\vec q_1
\right)^2 \right)^{\epsilon}~,
$$
\begin{equation}\label{528}
K_{4\perp\perp}^{\mu\nu} = \int_0^1d\beta\beta(1 - \beta)\left( \left(
(1 - \beta)\vec q_1^{\:\prime} + \beta\vec q_1 \right)^2 \right)
^{\epsilon}\frac{\left( (1 - \beta)q_1^{\prime} + \beta q_1 \right)_{\perp}
^{\mu}\left( (1 - \beta)q_1^{\prime} + \beta q_1 \right)_{\perp}^{\nu}}
{\left( (1 - \beta)q_1^{\prime} + \beta q_1 \right)_{\perp}^2}~,
\end{equation}
which appear in Eq.~(\ref{527}), cannot be calculated at arbitrary
$\epsilon$. 
We evaluate them in the Appendix in form of an expansion in $\epsilon$. Let us
note that the evident relation
\begin{equation}\label{529}
K_4 = K_{4\perp\mu}^{\mu\perp}~,
\end{equation}
is not very useful for us, because these two integrals should be calculated
with different accuracy (see factors in front of these integrals in
Eq.~(\ref{527})), so that we are forced to consider them separately. 

At this
point the calculation of the two-gluon contribution to the gluon impact 
factors 
defined in Eq.~(\ref{28}) is completed. The two-gluon contribution is given
by the relation (\ref{518}) with the color structures
${C_3}_{AA^{\prime}}^{cc^{\prime}}$ and ${C_4}_{AA^{\prime}}^{cc^{\prime}}$
shown in Eq.~(\ref{54}) and the coefficients $I_3$ and $I_4$ presented in 
Eqs.~(\ref{523}) and (\ref{527}), respectively. 

To obtain the impact factors 
$\Phi_{A^{\prime}A}^{({\cal R}, \nu)}(\vec q_1, \vec q; s_0)$ with the
$t$-channel state $\nu$ of the irreducible representation ${\cal R}$ of the
color group SU(N), it is enough to perform the convolution in Eq.~(\ref{27}).

\section{The check of the bootstrap}
\setcounter{equation}{0}

We have now all the contributions needed to check the bootstrap condition given
in Eq.~(\ref{24}). First of all, we must consider the gluon non-forward
impact factors in a generic state $a$ of the octet color representation in 
the $t$-channel. According to Eq.~(\ref{27}), this means that the four
one-loop order contributions to the unprojected gluon impact factors, due to 
one-gluon, quark-antiquark and two-gluon intermediate states given in
Eqs.~(\ref{313}), (\ref{46}) and (\ref{511}) respectively, and to the
counterterm given in Eq.~(\ref{516}) must be contracted with
\begin{equation}\label{61}
\langle cc^{\prime}|\hat{\cal P}_8|a\rangle =  \frac{iT_{cc^{\prime}}^a}
{\sqrt{N}}~.
\end{equation}
Since the projection on the octet color state $a$ gives
$$
i\sqrt{N}\langle cc^{\prime}|\hat{\cal P}_8|a\rangle\left(T^{c^{\prime}}T^c
\right)_{A^{\prime}A} = \frac{N}{2}T_{A^{\prime}A}^a~,\ \ \ \ \
i\sqrt{N}\langle cc^{\prime}|\hat{\cal P}_8|a\rangle{C_1}_{AA^{\prime}}
^{cc^{\prime}} = \frac{N}{4}T_{A^{\prime}A}^a~,
$$
\begin{equation}\label{62}
i\sqrt{N}\langle cc^{\prime}|\hat{\cal P}_8|a\rangle{C_3}_{AA^{\prime}}
^{cc^{\prime}} = \frac{N^2}{4}T_{A^{\prime}A}^a~,\ \ \ \ \ i\sqrt{N}\langle cc
^{\prime}|\hat{\cal P}_8|a\rangle{C_2}_{AA^{\prime}}^{cc^{\prime}} =
i\sqrt{N}\langle cc^{\prime}|\hat{\cal P}_8|a\rangle{C_4}_{AA^{\prime}}^{cc
^{\prime}} = 0~,
\end{equation}
we obtain for the octet gluon impact factors 
$$
i\sqrt{N}\Phi_{A^{\prime}A}^{(8, a)(1)}(\vec q_1, \vec q; s_0) = 
i\sqrt{N}\langle cc^{\prime}|\hat{\cal P}_8|a\rangle\Phi_{AA^{\prime}}
^{cc^{\prime}(1)}(\vec q_1, \vec q; s_0) = - \frac{1}{4}T_{A^{\prime}A}^ag^2N
\left( e_{A^{\prime}\perp}^*e_{A\perp} \right)\omega^{(1)}(- \vec q^{\:2})
$$
$$
\times\ln\left( \frac{s_0}{\vec q^{\:2}} \right) + T_{A^{\prime}A}^ag^4N^2
\frac{\Gamma(1 - \epsilon)}{2(4\pi)^{2 + \epsilon}}\frac
{\Gamma^2(1 + \epsilon)}{\Gamma(1 + 2\epsilon)}e_{A^{\prime}\mu}^{*\perp}
e_{A\nu}^{\perp}\biggl[ g_{\perp\perp}^{\mu\nu}\frac{1}{\epsilon}\biggl(
\frac{1}{\epsilon} - \frac{4}{(1 + 2\epsilon)}+ \frac{1}{(1 + \epsilon)
(3 + 2\epsilon)}
$$
$$
+ 2\psi(1) - 2\psi(1 + 2\epsilon) \biggr)\left( \vec q^{\:2}
\right)^{\epsilon} - \frac{1}{(1 + \epsilon)(1 + 2\epsilon)(3 + 2\epsilon)}
T_{\perp\perp}^{\mu\nu}(q_{\perp})\left( \vec q^{\:2} \right)
^{\epsilon} \biggr] + T_{A^{\prime}A}^ag^4N
$$
$$
\times\frac{\Gamma(1 - \epsilon)}{4(1 + \epsilon)(4\pi)^{2 +
\epsilon}}e_{A^{\prime}\mu}^{*\perp}e_{A\nu}^{\perp}\biggl[ g_{\perp\perp}
^{\mu\nu}F_1^{(+)}(\vec q^{\:2}) + 4T_{\perp\perp}^{\mu\nu}(q_{\perp})F_1^{(-)}
(\vec q^{\:2}) \biggr] + T_{A^{\prime}A}^ag^4N^2\frac{\Gamma
(1 - \epsilon)}{2(4\pi)^{2 + \epsilon}}
$$
$$
\times\frac{\Gamma^2(1 + \epsilon)}{\Gamma(1 + 2\epsilon)}\left( e_{A^{\prime}
\perp}^*e_{A\perp} \right)\biggl[ \frac{1}{2\epsilon}\biggl( \frac{1}
{\epsilon} + \frac{(11 + 7\epsilon)}{(1 + 2\epsilon)(3 + 2\epsilon)} + 2\psi
(1 + 2\epsilon) - 2\psi(1 + \epsilon) \biggr)
$$
$$
\times\biggl( \left( \vec q_1^{\:\prime\:2} \right)^{\epsilon}
+ \left( \vec q_1^{\:2} \right)^{\epsilon} \biggr) + K_1 \biggr]
- T_{A^{\prime}A}^ag^4N\frac{\Gamma(1 - \epsilon)}{(4\pi)^{2 +
\epsilon}}\left( e_{A^{\prime}\perp}^*e_{A\perp} \right)\frac{1}{\epsilon}
$$
\begin{equation}\label{63}
\times\sum_f\int_0^1dxx(1 - x)\biggl[ \left( m_f^2 + x(1 - x)\vec q_1
^{\:\prime\:2} \right)^{\epsilon} + \left( m_f^2 + x(1 - x)\vec
q_1^{\:2} \right)^{\epsilon} \biggr]~,
\end{equation}
the one-loop Reggeized gluon trajectory $\omega^{(1)}$ being given in 
Eq.~(\ref{311}), the tensor $T_{\perp\perp}^{\mu\nu}$ in Eq.~(\ref{35}), 
the functions $F_1^{(\pm)}$ in Eq.~(\ref{39}) and the integral $K_1$ in 
Eq.~(\ref{521}). Using the integral representation (\ref{311}) for the
one-loop gluon Regge trajectory, we get for the L.H.S. of the bootstrap
condition (\ref{24})
$$
{\mbox{L.H.S.}} = - \frac{1}{2}T_{A^{\prime}A}^ag
\left( e_{A^{\prime}\perp}^*e_{A\perp}
\right)\left( \omega^{(1)}(- \vec q^{\:2}) \right)^2\ln\left( \frac{s_0}
{\vec q^{\:2}} \right) + T_{A^{\prime}A}^ag^3N\frac{\Gamma(1 -
\epsilon)}{(4\pi)^{2 + \epsilon}}\frac{\Gamma^2(1 + \epsilon)}{\Gamma(1 +
2\epsilon)}
$$
$$
\times e_{A^{\prime}\mu}^{*\perp}e_{A\nu}^{\perp}\omega^{(1)}(- \vec q^{\:2})
\biggl[ g_{\perp\perp}^{\mu\nu}\frac{1}{\epsilon}\biggl( \frac{1}{\epsilon}
- \frac{4}{(1 + 2\epsilon)} + \frac{1}{(1 + \epsilon)(3 + 2\epsilon)} + 2\psi
(1) - 2\psi(1 + 2\epsilon) \biggr)\left( \vec q^{\:2} \right)
^{\epsilon}
$$
$$
- \frac{1}{(1 + \epsilon)(1 + 2\epsilon)(3 + 2\epsilon)}T_{\perp\perp}
^{\mu\nu}(q_{\perp})\left( \vec q^{\:2} \right)^{\epsilon}
\biggr] + T_{A^{\prime}A}^ag^3\frac{\Gamma(1 - \epsilon)}{2(1
+ \epsilon)(4\pi)^{2 + \epsilon}}e_{A^{\prime}\mu}^{*\perp}e_{A\nu}^{\perp}
\omega^{(1)}(- \vec q^{\:2})
$$
$$
\times\biggl[ g_{\perp\perp}^{\mu\nu}F_1^{(+)}(\vec q^{\:2}) + 4T_{\perp\perp}
^{\mu\nu}(q_{\perp})F_1^{(-)}(\vec q^{\:2}) \biggr] - T_{A^{\prime}A}^ag^5N^2
\frac{\Gamma(1 - \epsilon)}{2(4\pi)^{2 + \epsilon}}\frac{\Gamma
^2(1 + \epsilon)}{\Gamma(1 + 2\epsilon)}\left( e_{A^{\prime}\perp}^*e_{A\perp}
\right)
$$
$$
\times\int\frac{d^{D-2}q_1}{(2\pi)^{D-1}}\frac{\vec q^{\:2}}{\vec q_1^{\:2}(\vec
q_1 - \vec q)^2}\biggl[ \frac{1}{\epsilon}\biggl( \frac{1}{\epsilon} + \frac
{(11 + 7\epsilon)}{(1 + 2\epsilon)(3 + 2\epsilon)} + 2\psi(1 + 2\epsilon) -
2\psi(1 + \epsilon) \biggr)\left( \vec q_1^{\:2} \right)
^{\epsilon} + K_1 \biggr]
$$
$$
+ 2T_{A^{\prime}A}^ag^5N\frac{\Gamma(1 - \epsilon)}{(4\pi)
^{2 + \epsilon}}\left( e_{A^{\prime}\perp}^*e_{A\perp} \right)\frac{1}
{\epsilon}\int\frac{d^{D-2}q_1}{(2\pi)^{D-1}}\frac{\vec q^{\:2}}{\vec q_1^{\:2}
(\vec q_1 - \vec q)^2}
$$
\begin{equation}\label{64}
\times\sum_f\int_0^1dxx(1 - x)\left( m_f^2 + x(1 - x)\vec q_1^{\:2}
 \right)^{\epsilon}~.
\end{equation}
The analogous expression for the R.H.S. of Eq.~(\ref{24}) takes the form
$$
{\mbox{R.H.S.}} = - \frac{1}{2}T_{A^{\prime}A}^ag
\left( e_{A^{\prime}\perp}^*e_{A\perp}
\right)\omega^{(2)}(- \vec q^{\:2}) - \frac{1}{2}T_{A^{\prime}A}^ag\left(
e_{A^{\prime}\perp}^*e_{A\perp} \right)\left( \omega^{(1)}(- \vec q^{\:2})
\right)^2\ln\left( \frac{s_0}{\vec q^{\:2}} \right)
$$
$$
- T_{A^{\prime}A}^ag^3N\frac{\Gamma(1 - \epsilon)}{(4\pi)^{2 +
\epsilon}}\frac{\Gamma^2(1 + \epsilon)}{\Gamma(1 + 2\epsilon)}e_{A^{\prime}\mu}
^{*\perp}e_{A\nu}^{\perp}\omega^{(1)}(- \vec q^{\:2})\biggl[ g_{\perp\perp}
^{\mu\nu}\frac{1}{\epsilon}\biggl( - \frac{2}{\epsilon} + \frac{9(1 + \epsilon)
^2 + 2}{2(1 + \epsilon)(1 + 2\epsilon)(3 + 2\epsilon)}
$$
$$
+ 2\psi(1 + \epsilon) - \psi(1 - \epsilon) - \psi(1) \biggr)\left( \vec
q^{\:2} \right)^{\epsilon} + \frac{1}{(1 + \epsilon)(1 + 2\epsilon)(3
+ 2\epsilon)}T_{\perp\perp}^{\mu\nu}(q_{\perp})\left( \vec q^{\:2}
\right)^{\epsilon} \biggr]
$$
$$
+ T_{A^{\prime}A}^ag^3\frac{\Gamma(1 - \epsilon)}{2(1
+ \epsilon)(4\pi)^{2 + \epsilon}}e_{A^{\prime}\mu}^{*\perp}e_{A\nu}^{\perp}
\omega^{(1)}(- \vec q^{\:2})\biggl[ g_{\perp\perp}^{\mu\nu}F_1^{(+)}(\vec
q^{\:2}) + 4T_{\perp\perp}^{\mu\nu}(q_{\perp})F_1^{(-)}(\vec q^{\:2}) \biggr]
$$
\begin{equation}\label{65}
- 2T_{A^{\prime}A}^ag^3\frac{\Gamma(1 - \epsilon)}{(4\pi)
^{2 + \epsilon}}\left( e_{A^{\prime}\perp}^*e_{A\perp} \right)\frac{1}
{\epsilon}\omega^{(1)}(- \vec q^{\:2})\sum_f\int_0^1dxx(1 - x)\left( m_f^2
+ x(1 - x)\vec q^{\:2} \right)^{\epsilon}~,
\end{equation}
where $\omega^{(2)}$ is the two-loop correction to the Reggeized gluon 
trajectory and we have used Eqs.~(\ref{32})-(\ref{35}), (\ref{38}) and 
(\ref{39}) which give the gluon-gluon-Reggeon effective vertex with one-loop 
accuracy.

We see that the helicity non-conserving parts (the terms with the tensor
$T_{\perp\perp}^{\mu\nu}$) are completely cancelled in both L.H.S. and
R.H.S. of the bootstrap condition.  This fact  is very important, because
up to now the possibility to present in the Regge form the helicity 
non-conserving part 
of the elastic scattering amplitude with gluon color quantum numbers 
in the $t$-channel was not checked anywhere.

As for the helicity conserving part, the bootstrap condition for it 
is not quite independent from the calculation of the 
two-loop correction to the gluon trajectory~\cite{4}, since it was 
performed  
assuming that the gluon Reggeization holds, by comparison of 
the $s$-channel 
discontinuity dictated by the Regge form~(\ref{Ao}) with that 
calculated from the 
unitarity. Therefore, the check of the bootstrap condition 
for the helicity conserving part gives more
a test of correctness of all the calculations involved in the 
determination of the 
trajectory and of the impact factors. 

We notice that there are cancellations
between the terms in Eq.~(\ref{64}) and (\ref{65}) containing the factors 
$\ln(s_0/\vec q^{\:2})$ and the functions
$F_1^{(\pm)}$, respectively. Then we arrive at the following equality,
which must be valid for the bootstrap:
$$
\omega^{(2)}(- \vec q^{\:2}) = \omega^{(2)(G)}(- \vec q^{\:2}) + \omega^{(2)(Q)}
(- \vec q^{\:2}) = g^4N^2\frac{\Gamma(1 - \epsilon)}{2(4\pi)^{2 +
\epsilon}}\frac{\Gamma^2(1 + \epsilon)}{\Gamma(1 + 2\epsilon)}\int\frac
{d^{D-2}q_1}{(2\pi)^{D-1}}
$$
$$
\times\frac{\vec q^{\:2}}{\vec q_1^{\:2}(\vec q_1 - \vec q)^2}\biggl[ \frac{1}
{\epsilon}\biggl( \frac{1}{\epsilon} + \frac{(11 + 7\epsilon)}{(1 + 2\epsilon)
(3 + 2\epsilon)} + 2\psi(1 + 2\epsilon) - 2\psi(1 + \epsilon) \biggr)\biggl(
2\left( \vec q_1^{\:2} \right)^{\epsilon} - \left( \vec
q^{\:2} \right)^{\epsilon} \biggr)
$$
$$
- \frac{1}{\epsilon}\biggl( \frac{1}{\epsilon} + 2\psi(1 + 2\epsilon) - 2\psi
(1 + \epsilon) + 2\psi(1 - \epsilon) - 2\psi(1) \biggr)\left( \vec
q^{\:2} \right)^{\epsilon} + 2K_1 \biggr] + 2g^4N\frac{
\Gamma(1 - \epsilon)}{(4\pi)^{2 + \epsilon}}\frac{1}{\epsilon}
$$
$$
\times\int\frac{d^{D-2}q_1}{(2\pi)^{D-1}}\frac{\vec q^{\:2}}{\vec q_1^{\:2}
(\vec q_1 - \vec q)^2}\sum_f\int_0^1dxx(1 - x)
$$
\begin{equation}\label{66}
\times\biggl[ \left( m_f^2 + x(1 - x)\vec q^{\:2} \right)
^{\epsilon} - 2\left( m_f^2 + x(1 - x)\vec q_1^{\:2} \right)
^{\epsilon} \biggr]~.
\end{equation}
Here $\omega^{(2)(G)}$ and $\omega^{(2)(Q)}$ are the gluon and quark 
contributions to the two-loop correction to the gluon Regge trajectory 
$\omega^{(2)}$; they are known and can be found, for example, in
Ref.~\cite{14}. 
The cancellation of the quark contribution in both sides of Eq.~(\ref{66}) is 
evident if one compares this equation with the expression (71) of
Ref.~\cite{14} for $\omega^{(2)(Q)}$. Let us check the
cancellation of the gluon contribution. To do this, consider the integral
representation of Ref.~\cite{14} for the gluon part $\omega^{(2)(G)}$ of 
the two-loop correction to the gluon trajectory:
$$
\omega^{(2)(G)}(- \vec q^{\:2}) = \frac{g^4N^2}{2}\int\frac{d^{D-2}q_1}
{(2\pi)^{D-1}}\frac{d^{D-2}q_2}{(2\pi)^{D-1}}\frac{\vec q^{\:2}}{\vec q_1^{\:2}
\vec q_2^{\:2}}\biggl[ \frac{\vec q^{\:2}}{2(\vec q_1 - \vec q)^2(\vec q_2 -
\vec q)^2}\ln\left( \frac{\vec q^{\:2}}{(\vec q_1 - \vec q_2)^2} \right)
$$
$$
- \frac{1}{(\vec q_1 + \vec q_2 - \vec q)^2}\ln\left( \frac{(\vec q_1 -
\vec q)^2}{\vec q_1^{\:2}} \right) + \biggl( \frac{- \vec q^{\:2}}{2(\vec q_1 -
\vec q)^2(\vec q_2 - \vec q)^2} + \frac{1}{(\vec q_1 + \vec q_2 - \vec q)^2}
\biggr)
$$
\begin{equation}\label{67}
\times\biggl( \frac{1}{\epsilon} + \frac{(11 + 7\epsilon)}{2(1 + 2\epsilon)
(3 + 2\epsilon)} + 2\psi(1 + 2\epsilon) - 2\psi(1 + \epsilon) + \psi(1 -
\epsilon) - \psi(1) \biggr) \biggr]~.
\end{equation}
It can be written in the form
$$
\omega^{(2)(G)}(- \vec q^{\:2}) = \frac{g^4N^2}{2}\int\frac{d^{D-2}q_1}
{(2\pi)^{D-1}}\frac{\vec q^{\:2}}{\vec q_1^{\:2}(\vec q_1 - \vec q)^2}\biggl[
\frac{1}{2}\int\frac{d^{D-2}k}{(2\pi)^{D-1}}\ln\left( \frac{\vec q^{\:2}}
{\vec k^{\:2}} \right)\frac{\vec q^{\:2}}{(\vec k - \vec q_1^{\:\prime})^2
(\vec k - \vec q_1)^2} -
$$
$$
\int\frac{d^{D-2}k}{(2\pi)^{D-1}}\ln\left( \frac{\vec q_1^{\:2}}{\vec k^{\:2}}
\right)\frac{\vec q_1^{\:2}}{\vec k^{\:2}(\vec k - \vec q_1)^2} + \frac{1}{2}
\biggl( 2\int\frac{d^{D-2}k}{(2\pi)^{D-1}}\frac{\vec q_1^{\:2}}{\vec k^{\:2}
(\vec k - \vec q_1)^2} - \int\frac{d^{D-2}k}{(2\pi)^{D-1}}\frac{\vec q^{\:2}}
{\vec k^{\:2}(\vec k - \vec q)^2} \biggr)
$$
\begin{equation}\label{68}
\times\biggl(\frac{1}{\epsilon} + \frac{(11 + 7\epsilon)}{2(1 + 2\epsilon)
(3 + 2\epsilon)} + 2\psi(1 + 2\epsilon) - 2\psi(1 + \epsilon) + \psi(1 -
\epsilon) - \psi(1) \biggr) \biggr]~.
\end{equation}
Let us now express the first integral into square brackets of this equation
through $K_1$, given in Eq.~(\ref{521}), and calculate exactly the other 
integrals, making
use of the relations (\ref{311}) and (\ref{520}). So doing, 
we get just the 
first term in the R.H.S. of Eq.~(\ref{66}),
what proves that the bootstrap condition (\ref{24}) is satisfied
as far as the gluon NLA impact factors are concerned. 

To summarize, we have verified the fulfillment of the bootstrap conditions for 
the helicity non-conserving and conserving parts, separately.
We stress that, while for the helicity non-conserving part 
of Eq.~(\ref{24}) it gives the first check of the possibility to 
present it in the Regge form, for the helicity conserving part it represents
also a check of the previous result of Ref.~\cite{4} for the two-loop
Reggeized gluon trajectory $\omega^{(2)}$. So, at least the integral 
representation for $\omega^{(2)}$ obtained there is correct. But since there is 
also an independent check~\cite{BRN98} of the integrated result of 
Ref.~\cite{5} for $\omega^{(2)}$, the final result of that paper is fully 
verified.

\section{Gluon impact factors in massless QCD}
\setcounter{equation}{0}

In this section we calculate explicitly the gluon NLA non-forward impact 
factors (\ref{28}), restricting ourselves to the case of massless quarks. 
In this case most integrations can be carried out exactly at arbitrary 
$\epsilon$ and 
there are only few integrals which we are forced to calculate as expansion
in $\epsilon$.
The Born contribution to the impact factor, which we present here for the
sake of completeness, is not changed, of course; it is (see Eq.~(\ref{312}))
\begin{equation}\label{71}
\Phi_{AA^{\prime}}^{cc^{\prime}(B)}(\vec q_1, \vec q) = - g^2\left(
T^{c^{\prime}}T^c \right)_{A^{\prime}A}e_{A^{\prime}\perp}^{*\mu}
e_{A\perp}^{\nu}g_{\mu\nu}^{\perp\perp}~.
\end{equation}
The integration in Eq.~(\ref{313}), as well as the integrations in 
Eqs.~(\ref{39}), which give the functions $F_1^{(\pm)}$ can be easily 
done, leading to
$$
\sum_f\frac{2(1 + \epsilon)}{\epsilon}\int_0^1dxx(1 - x)\left( m_f^2
+ x(1 - x)\vec v^{\:2} \right)^{\epsilon}\biggl|_{m_f = 0}
$$
\begin{equation}\label{72}
= \frac{n_f\Gamma^2(1 + \epsilon)}{\Gamma(1 + 2\epsilon)}\frac{(1 + \epsilon)
^2}{\epsilon(1 + 2\epsilon)(3 + 2\epsilon)}\left( \vec v^{\:2}
\right)^{\epsilon}~,
\end{equation}
$$
F_1^{(+)}(\vec v^{\:2})\biggl|_{m_f = 0} = \frac{n_f\Gamma^2(1 + \epsilon)}
{\Gamma(1 + 2\epsilon)}\frac{2}{\epsilon}\biggl( \frac{(1 + \epsilon)}{(1 +
2\epsilon)} - \frac{1}{(1 + \epsilon)(3 + 2\epsilon)} \biggr)
\left( \vec v^{\:2} \right)^{\epsilon}~,
$$
\begin{equation}\label{73}
F_1^{(-)}(\vec v^{\:2})\biggl|_{m_f = 0} = \frac{n_f\Gamma^2(1 + \epsilon)}
{\Gamma(1 + 2\epsilon)}\frac{1}{2(1 + \epsilon)(1 + 2\epsilon)(3 + 2\epsilon)}
\left( \vec v^{\:2} \right)^{\epsilon}~,
\end{equation}
where $n_f$ is the number of light quark flavors. Then from 
Eq.~(\ref{313}), with the help of last three relations and of Eq.~(\ref{35}), 
we get the one-gluon contribution to the one-loop correction for the gluon impact 
factors (\ref{28}) in the massless quark case:
$$
\Phi_{AA^{\prime}}^{cc^{\prime}(1)\{G\}}(\vec q_1, \vec q; s_0)\biggl( g^4N^2
\frac{\Gamma(1 - \epsilon)}{(4\pi)^{2 + \epsilon}}\frac
{\Gamma^2(1 + \epsilon)}{\Gamma(1 + 2\epsilon)} \biggr)^{-1} = \frac{1}{N}
\left( T^{c^{\prime}}T^c \right)_{A^{\prime}A}e_{A^{\prime}\mu}^{*\perp}
e_{A\nu}^{\perp}
$$
$$
\times\biggl[ g_{\perp\perp}^{\mu\nu}\frac{1}{\epsilon}\biggl( \ln\left( \frac
{s_0}{\vec q_1^{\:\prime\:2}} \right)\left( \vec q_1^{\:\prime\:2}
\right)^{\epsilon} + \ln\left( \frac{s_0}{\vec q_1^{\:2}} \right)
\left( \vec q_1^{\:2} \right)^{\epsilon} \biggr) + g_{\perp\perp}^{\mu\nu}
\frac{1}{\epsilon}\biggl( \frac{2}{\epsilon} - \frac{(11 + 9\epsilon)}{2(1 +
2\epsilon)(3 + 2\epsilon)}
$$
$$
+ \frac{n_f}{N}\frac{(1 + \epsilon)(2 + \epsilon) - 1}{(1 + \epsilon)(1 +
2\epsilon)(3 + 2\epsilon)} + \psi(1) + \psi(1 - \epsilon) - 2\psi(1 + \epsilon)
\biggr)\biggl( \left( \vec q_1^{\:\prime\:2} \right)^{\epsilon} +
\left( \vec q_1^{\:2} \right)^{\epsilon} \biggr)
$$
\begin{equation}\label{74}
+ \frac{2}{(1 + \epsilon)(1 + 2\epsilon)(3 + 2\epsilon)}\left( 1 + \epsilon -
\frac{n_f}{N} \right)\biggl( \frac{q_{1\perp}^{\prime\:\mu}q_{1\perp}
^{\prime\:\nu}}{q_{1\perp}^{\prime\:2}}\left( \vec q_1^{\:\prime\:2}
\right)^{\epsilon} + \frac{q_{1\perp}^{\mu}q_{1\perp}^{\nu}}{q_{1\perp}^2}
\left( \vec q_1^{\:2} \right)^{\epsilon} \biggr) \biggr]~.
\end{equation}

For the case of the quark-antiquark intermediate state contribution, all
integrations in Eq.~(\ref{421}) (see for instance Eqs.~(\ref{73})) and the
integration over the variable $x$ in Eq.~(\ref{424}) can be easily
performed in exact form. As for integration over the variable $\beta$ in 
(\ref{424}), it cannot be completely carried out for arbitrary $\epsilon$. 
With help of the Eqs.~(\ref{35}), (\ref{47}), (\ref{413}), (\ref{421}), 
(\ref{424}) and (\ref{73}) we obtain
$$
\Phi_{AA^{\prime}}^{cc^{\prime}(1)\{q\bar q\}}(\vec q_1, \vec q)\biggl( g^4N^2
\frac{\Gamma(1 - \epsilon)}{(4\pi)^{2 + \epsilon}}\frac
{\Gamma^2(1 + \epsilon)}{\Gamma(1 + 2\epsilon)} \biggr)^{-1} = \frac{2}{N}
tr\left( t^At^ct^{c^{\prime}}t^{A^{\prime}} + t^At^{A^{\prime}}t^{c^{\prime}}
t^c \right)e_{A^{\prime}\mu}^{*\perp}e_{A\nu}^{\perp}\frac{n_f}{N}
$$
$$
\times\biggl[ - g_{\perp\perp}^{\mu\nu}\frac{2(1 + \epsilon)^2 + \epsilon}
{\epsilon(1 + \epsilon)(1 + 2\epsilon)(3 + 2\epsilon)}\biggl
( \left( \vec q_1^{\:\prime\:2} \right)^{\epsilon} + 
\left( \vec q_1^{\:2} \right)^{\epsilon} - \left( \vec q^{\:2} \right)
^{\epsilon} \biggr) + \frac{2}{(1 + \epsilon)(1 + 2\epsilon)(3 + 2\epsilon)}
$$
$$
\times\biggl( \frac{q_{1\perp}^{\prime\:\mu}q_{1\perp}^{\prime\:\nu}}{q_{1\perp}
^{\prime\:2}}\left( \vec q_1^{\:\prime\:2} \right)^{\epsilon} +
\frac{q_{1\perp}^{\mu}q_{1\perp}^{\nu}}{q_{1\perp}^2}\left( \vec q_1
^{\:2} \right)^{\epsilon} - \frac{q_{\perp}^{\mu}q_{\perp}^{\nu}}
{q_{\perp}^2}\left( \vec q^{\:2} \right)^{\epsilon} \biggr)
\biggr] + \frac{2}{N}tr\left( t^At^ct^{A^{\prime}}t^{c^{\prime}} + t^At
^{c^{\prime}}t^{A^{\prime}}t^c \right)e_{A^{\prime}\mu}^{*\perp}e_{A\nu}
^{\perp}
$$
$$
\times\frac{n_f}{N}\biggl[ g_{\perp\perp}^{\mu\nu}\frac{(2 + \epsilon)}
{\epsilon(1 + \epsilon)(3 + 2\epsilon)}\biggl( \left( \vec q_1^{\:\prime
\:2} \right)^{\epsilon} + \left( \vec q_1^{\:2} \right)
^{\epsilon} \biggr) - \frac{2}{(1 + \epsilon)(1 + 2\epsilon)(3 + 2\epsilon)}
\biggl( \frac{q_{1\perp}^{\prime\:\mu}q_{1\perp}^{\prime\:\nu}}{q_{1\perp}^{\prime
\:2}}\left( \vec q_1^{\:\prime\:2} \right)^{\epsilon}
$$
\begin{equation}\label{75}
+ \frac{q_{1\perp}^{\mu}q_{1\perp}^{\nu}}{q_{1\perp}^2}\left( \vec q_1
^{\:2} \right)^{\epsilon} \biggr) - g_{\perp\perp}^{\mu\nu}\biggl(
\frac{1}{\epsilon}K_3 - \frac{2}{\epsilon(1 + 2\epsilon)}K_4 \biggr) + \frac
{4}{(1 + 2\epsilon)}K_{4\perp\perp}^{\mu\nu} \biggr]~,
\end{equation}
with the integrals $K_3$, $K_4$ and $K_{4\perp\perp}^{\mu\nu}$ presented in the
Eqs.~(\ref{528}).

As for the NLA two-gluon contribution to the gluon impact factors for the 
non-forward scattering, found in Section 5,  it does not change in the massless
quark case. For completeness we write down its expression below with help of
Eqs.~(\ref{35}), (\ref{54}), (\ref{518}), (\ref{523}), (\ref{524}) and 
(\ref{527}):
$$
\Phi_{AA^{\prime}}^{cc^{\prime}(1)\{GG\}}(\vec q_1, \vec q; s_0)\biggl( g^4N^2
\frac{\Gamma(1 - \epsilon)}{(4\pi)^{2 + \epsilon}}\frac
{\Gamma^2(1 + \epsilon)}{\Gamma(1 + 2\epsilon)} \biggr)^{-1} = - \frac{2}{N^2}
\left( T^{c_1^{\prime}}T^{c_1} \right)_{A^{\prime}A}\left( T^{c_1^{\prime}}
T^{c_1} \right)_{c^{\prime}c}
$$
$$
\times e_{A^{\prime}\mu}^{*\perp}e_{A\nu}^{\perp}\biggl[ g_{\perp\perp}
^{\mu\nu}\frac{1}{\epsilon}\biggl( \ln\left( \frac{s_0}{\vec q_1^{\:\prime\:2}}
\right)\left( \vec q_1^{\:\prime\:2} \right)^{\epsilon} + \ln\left(
\frac{s_0}{\vec q_1^{\:2}} \right)\left( \vec q_1^{\:2} \right)
^{\epsilon} - \ln\left( \frac{s_0}{\vec q^{\:2}} \right)\left( \vec q
^{\:2} \right)^{\epsilon} \biggr) + g_{\perp\perp}^{\mu\nu}\frac{1}
{\epsilon}\biggl( \frac{3}{2\epsilon}
$$
$$
- \frac{(11 + 8\epsilon)}{(1 + 2\epsilon)(3 + 2\epsilon)} - \psi(1 + 2\epsilon)
- \psi(1 + \epsilon) + \psi(1 - \epsilon) + \psi(1) \biggr)\biggl( 
\left( \vec q_1^{\:\prime\:2} \right)^{\epsilon} + \left( \vec q_1^{\:2} 
\right)^{\epsilon} - \left( \vec q^{\:2} \right)
^{\epsilon} \biggr)
$$
$$
+ g_{\perp\perp}^{\mu\nu}\frac{1}{2\epsilon}\biggl( \frac{1}{\epsilon} + 2\psi
(1 + 2\epsilon) - 2\psi(1 + \epsilon) + 2\psi(1 - \epsilon) - 2\psi(1) \biggr)
\left( \vec q^{\:2} \right)^{\epsilon} - g_{\perp\perp}^{\mu\nu}
K_1 + \frac{2}{(1 + 2\epsilon)(3 + 2\epsilon)}
$$
$$
\times\biggl( \frac{q_{1\perp}^{\prime\:\mu}q_{1\perp}^{\prime\:\nu}}{q_{1\perp}
^{\prime\:2}}\left( \vec q_1^{\:\prime\:2} \right)^{\epsilon} +
\frac{q_{1\perp}^{\mu}q_{1\perp}^{\nu}}{q_{1\perp}^2}\left( \vec q_1
^{\:2} \right)^{\epsilon} - \frac{q_{\perp}^{\mu}q_{\perp}^{\nu}}
{q_{\perp}^2}\left( \vec q^{\:2} \right)^{\epsilon} \biggr)
\biggr] - \frac{2}{N^2}\left( T^{c_1^{\prime}}T^{c_1} \right)_{c^{\prime}A}
\left( T^{c_1^{\prime}}T^{c_1} \right)_{A^{\prime}c}e_{A^{\prime}\mu}^{*\perp}
e_{A\nu}^{\perp}
$$
$$
\times\biggl[ g_{\perp\perp}^{\mu\nu}\frac{(11 + 8\epsilon)}{\epsilon(1 +
2\epsilon)(3 + 2\epsilon)}\biggl( \left( \vec q_1^{\:\prime\:2}
\right)^{\epsilon} + \left( \vec q_1^{\:2} \right)^{\epsilon}
\biggr) - \frac{2}{(1 + 2\epsilon)(3 + 2\epsilon)}\biggl( \frac{q_{1\perp}
^{\prime\:\mu}q_{1\perp}^{\prime\:\nu}}{q_{1\perp}^{\prime\:2}}\left( \vec
q_1^{\:\prime\:2} \right)^{\epsilon} + \frac{q_{1\perp}^{\mu}q_{1\perp}
^{\nu}}{q_{1\perp}^2}
$$
\begin{equation}\label{76}
\times\left( \vec q_1^{\:2} \right)^{\epsilon} \biggr) +
g_{\perp\perp}^{\mu\nu}\biggl( \frac{2}{\epsilon}K_2 - \frac{4}{\epsilon}K_3
+ \frac{2(1 + \epsilon)}{\epsilon(1 + 2\epsilon)}K_4 \biggr) + \frac{4(1 +
\epsilon)}{(1 + 2\epsilon)}K_{4\perp\perp}^{\mu\nu} \biggr]~,
\end{equation}
with the integrals $K_1$ and $K_2$ defined by Eqs.~(\ref{521}) and (\ref{528}),
respectively. So we see that, to obtain the gluon impact factors 
(\ref{28}) in the massless quark case, 
we should calculate the integrals $K_1$ - $K_4$ and $K_{4\perp\perp}^{\mu\nu}$.
This calculation can be done only in the form of an expansion in $\epsilon$,
as it was already mentioned above, and we perform it in the Appendix. In order to
understand what is the accuracy according to which the integrals must be 
calculated, we notice that further applications of the non-forward impact 
factors imply a subsequent integration over $\vec q_1$ in the form like
that in the bootstrap condition (\ref{24}) (see Ref.~\cite{12}). In this 
subsequent integration the integrand is singular in the
regions of  $\vec q_1 \rightarrow 0$ and $\vec q_1^{\:\prime} = \vec q_1
- \vec q \rightarrow 0$, so that in these limits one must have exact
expressions for the impact factors (or, at least, expressions which lead to 
an accuracy 
up to finite terms in the physical limit $\epsilon \rightarrow 0$ after the
integration over $\vec q_1$). An analogous situation was observed in the 
calculation of the forward BFKL equation kernel in Ref.~\cite{7} and
detailed explanations can be found there. In the regions where the integrand is 
non-singular, it is enough to know impact factors with accuracy up to terms of 
the type $\epsilon^0$. The above discussion, taking also into account the
coefficients in the integrals $K_1$ - $K_4$ and $K_{4\perp\perp}
^{\mu\nu}$ given by the expressions (\ref{75}) and (\ref{76}) for the
corresponding contributions to the impact factors, make clear what terms we 
should keep in the expansion in $\epsilon$ for each of these integrals.

For the case of the forward scattering, being
\begin{equation}\label{77}
\vec q = 0,\ \ \ \ \ \ \vec q_1^{\:\prime} = \vec q_1~,
\end{equation}
we have further simplifications, which lead to the possibility to calculate
the integrals $K_1$ - $K_4$ and $K_{4\perp\perp}^{\mu\nu}$ in exact form
without expansion in $\epsilon$; we find
$$
K_1 = 0~,\ \ \ \ \ \ K_2 = \left( - \frac{1}{\epsilon} + 2\psi(1 +
2\epsilon) - 2\psi(1) \right)\left( \vec q_1^{\:2} 
\right)^{\epsilon}~,
$$
\begin{equation}\label{78}
K_3 = \left( \vec q_1^{\:2} \right)^{\epsilon},\ \ \ \ K_4 = 
\frac{1}{6}\left( \vec q_1^{\:2} \right)^{\epsilon},\ \ \ \ 
K_{4\perp\perp}^{\mu\nu} =
\frac{1}{6}\frac{q_{1\perp}^{\mu}q_{1\perp}^{\nu}}
{q_{1\perp}^2}\left( \vec q_1^{\:2} \right)^{\epsilon}~.
\end{equation}
Let us stress that the results for these integrals shown in the Appendix
for the non-forward case do not have the correct asymptotics as in the forward
one and can be used only for $\vec q \neq 0$. We did not care in
the Appendix about it, because these two cases are well separated and also 
because the integrals for the forward scattering are calculated
exactly. For example, we consider here the expression for the forward gluon
impact factor with the singlet color state in the $t$-channel, putting
also the helicity $\lambda_{A^{\prime}}$ to be equal to $\lambda_A$ and
taking an average over this quantum number. Consequently, the Born
contribution reads
\begin{equation}\label{79}
\Phi_G^{(0)(B)} = g^2\sqrt{\frac{N^2}{N^2 - 1}}
\end{equation}
and the one-loop one takes the form
$$
\Phi_G^{(0)(1)}(\vec q_1; s_0) = \Phi_G^{(0)(B)}\omega^{(1)}(- \vec q_1^{\:2})
\biggl[ - \ln\left( \frac{s_0}{\vec q_1^{\:2}} \right) + \frac{11}{6} - \frac
{2\epsilon}{(1 + 2\epsilon)(3 + 2\epsilon)} + \psi(1)
$$
\begin{equation}\label{710}
- \psi(1 - \epsilon) + \left( \frac{11}{3} - \frac{2}{3}\frac{n_f}{N} \right)
\frac{3(1 + \epsilon)}{2(1 + 2\epsilon)(3 + 2\epsilon)} + \frac{n_f}{N^3}\frac
{(2 + 3\epsilon)}{6(1 + \epsilon)} \biggr]~,
\end{equation}
where the one-loop Reggeized gluon trajectory $\omega^{(1)}$ is defined by
Eq.~(\ref{311}). In these two last equations we also have used the relation
(see Eq.~(\ref{27}))
\begin{equation}\label{711}
\langle cc^{\prime} | \hat{\cal P}_0 | 0 \rangle = \frac{\delta_{cc^{\prime}}}
{\sqrt{N^2 - 1}}
\end{equation}
and have omitted the common color factor $\delta_{AA^{\prime}}$.

The gluon impact factor in the forward case ($t=0$ and color singlet in the 
$t$-channel) was considered in Refs.~\cite{16,17}. In Ref.~\cite{17}
it was calculated for massless quarks  with accuracy up to terms finite in 
the $\epsilon \rightarrow 0$ limit. Unfortunately, the comparison of our 
result~(\ref{710}) for this particular case  with the corresponding 
result of 
Ref.~\cite{17} is not straightforward 
because of the different definitions adopted. First of all, we have 
used up
to now a fixed energy scale $s_0$ independent of the virtualities of the
Reggeized gluons. The transition to the general case of any factorizable
scale $s_0 = \sqrt{f_1(\vec q_1, \vec q) f_2(\vec q_2, \vec q)}$ in
Eq.~(\ref{Ar}) can be made to the NLA accuracy without changing the
Green function by the substitution~\cite{11}
\[
\Phi_{A^{\prime}A}^{({\cal R},\nu)}(\vec q_1, \vec q; s_0) \longrightarrow 
\]
\begin{equation}
\longrightarrow
\Phi_{A^{\prime}A}^{({\cal R},\nu)}(\vec q_1, \vec q; s_0)
+ \frac{1}{2} \int \frac{d^{D-2} q_r }
{\vec{q}_r^{\:2} \vec{q}_r^{\:\prime \:2}}
\Phi_{A^{\prime}A}^{({\cal R},\nu)(B)}(\vec q_r, \vec q)
{\cal K}^{({\cal R})(B)}\left( \vec q_r,\vec q_1;\vec{q}\right) 
\ln\left(\frac{f_1(\vec q_r, \vec q)}{s_0}\right)\;,
\end{equation}
where ${\cal K}^{({\cal R})(B)}$ is the non-forward BFKL kernel
in the LLA~\cite{1,10,12}
\[
{\cal K}^{({\cal R})(B)}(\vec q_1, \vec q_2, \vec q) = 
\vec{q}_{1}^{\:2} \vec{q}_{1}^{\:\prime\:2}\delta^{\left(D-2\right) }
\left( \vec{q}_{1}-\vec{q}_{2}\right) \, \biggl( \omega^{(1)}(-\vec q_1^{\:2})
+ \omega^{(1)}(-\vec q_1^{\:\prime\:2})\biggr)
\]
\begin{equation}
+ \frac{g^2}{(2\pi)^{D-1}} c_{{\cal R}} 
\left( \frac{\vec q_1^{\:2} \vec q_2^{\:\prime\:2}+\vec q_1^{\:\prime\:2}
\vec q_2^{\:2}}{(\vec q_1- \vec q_2)^2}-\vec q^{\:2}\right)\;.
\end{equation}
The coefficient $c_{{\cal R}}$ is equal to $N$ for the singlet and to $N/2$ 
for the octet representation. 
In the particular case of the forward helicity conserving impact factor and
Regge-motivated scale $s_0=|\vec q_1| |\vec q_2|$ used in Ref.~\cite{9},
it changes our result (\ref{710}) to the expression
$$
\frac{\Phi_G^{(0)(1)}(\vec q_1; s_0)}
{\Phi_G^{(0)(B)}} \longrightarrow  \omega^{(1)}(- \vec
q_1^{\:2})\biggl[ \left( \frac{11}{6} - \frac{n_f}{3N} \right) + \left( \frac
{11}{6} + \frac{(2 + \epsilon)n_f}{6N} \right) - \frac{C_F \, n_f}{N^2}\left( \frac
{2}{3} + \frac{\epsilon}{3} \right)
$$
\begin{equation}\label{c1}
- \frac{\epsilon}{N}\left( N\left( \frac{67}{18} - \frac{\pi^2}{6}
\right) - \frac{5n_f}{9} \right) \biggr] + g^2N\vec q_1^{\:2}\frac{1}{(2\pi)^
{D-1}}\int\frac{d^{D-2}q_r}{\vec q_r^{\:2}(\vec q_r - \vec q_1)^2}\ln\left(
\frac{\vec q_r^{\:2}}{\vec q_1^{\:2}} \right) ~,
\end{equation}
where $C_F=(N^2-1)/(2N)$ and we have also made an expansion in $\epsilon$ in 
the integrated part
with accuracy up to terms finite in the limit $\epsilon \rightarrow 0$.
The scale $s_0=|\vec q_1| |\vec q_2|$ was adopted also in~\cite{17},
but the impact factors were defined there on the basis of their infrared
properties, so that additional operator factors $H_L$ and $H_R$ were 
introduced~\cite{CC98,17} in the Green function. Therefore, one has to compare our
result with changed scale (\ref{c1}) not simply with the expression (5.11) of
Ref.~\cite{17}, but with this expression plus the piece connected with the
operator $H$. To our opinion, there is misprint in the expression (3.12) of
Ref.~\cite{17} for this operator, which should contain the additional factor
$1/\Gamma(1 - \epsilon)$. If so, one can easily obtain from Ref.~\cite{17}, that 
the account of operator $H$ leads to 
$$
\frac{h_g^{(1)}(\vec q_1)}{h_g^{(0)}(\vec q_1)} \longrightarrow 
\omega^{(1)}(- \vec q_1^{\:2})
\biggl[ \left( \frac{11}{6} - \frac{n_f}{3N} \right) + \left( \frac{11}{6} +
\frac{(2 + \epsilon)n_f}{6N} \right) - \frac{C_F\, n_f}{N^2}\left( \frac{2}{3} +
\frac{\epsilon}{3} \right)
$$
$$
- \frac{\epsilon}{N}\left( N\left( \frac{67}{18}
- \frac{\pi^2}{6} \right) -
\frac{5n_f}{9} \right) \biggr]
$$
\begin{equation}\label{c2}
- g^2N\vec q_1^{\:2}\frac{1}{(2\pi)^{D-1}}\int\frac{d^{D-2}q_r}{\vec
q_r^{\:2}(\vec q_r - \vec q_1)^2}\ln\left( \frac{(\vec q_r - \vec q_1)^2}{\vec
q_r^{\:2}} \right)\theta\biggl( (\vec q_r - \vec q_1)^2 - \vec q_r^{\:2} \biggr)
,
\end{equation}
where $h_g^{(0)}$ and $h_g^{(1)}$ are notations of Ref.~\cite{17} for 
Born and one-loop parts of the forward helicity-conserving color singlet gluon
impact factor. Note, $\Phi_G^{(0)(B)}$ and $h_g^{(0)}$ have different
normalization, but comparing Eq.~(\ref{Ar}) of the present paper and Eq.~(2.1) of 
Ref.~\cite{17}, it is quite easy to see that the R.H.S. in
(\ref{c1}) and (\ref{c2}) must coincide. Therefore we should check the equality
$$
\int\frac{d^{D-2}q_r}{\vec q_r^{\:2}(\vec q_r - \vec q_1)^2}\biggl[ \ln\left(
\frac{\vec q_r^{\:2}}{\vec q_1^{\:2}} \right) - \ln\left( \frac{\vec q_r^{\:2}}
{(\vec q_r - \vec q_1)^2} \right)\theta\biggl( (\vec q_r - \vec q_1)^2 - \vec
q_r^{\:2} \biggr) \biggr] =
$$
\begin{equation}\label{c3}
2\int\frac{d^{D-2}q_r}{\vec q_r^{\:2}(\vec q_r - \vec q_1)^2}\ln\left(
\frac{\vec q_r^{\:2}}{\vec q_1^{\:2}} \right)\theta\biggl( \vec q_r^{\:2} -
(\vec q_r - \vec q_1)^2 \biggr) = 0~,
\end{equation}
which must be satisfied with accuracy up to terms non-vanishing in the limit
$\epsilon = 0$. As it is evident from second line of Eq.~(\ref{c3}), the
expression there has no poles in $\epsilon$, therefore it is enough to check
that
$$
I = \int\frac{d^2q_r}{\vec q_r^{\:2}(\vec q_r - \vec q_1)^2}\ln\left(
\frac{\vec q_r^{\:2}}{\vec q_1^{\:2}} \right)\theta\biggl( \vec q_r^{\:2} -
(\vec q_r - \vec q_1)^2 \biggr) =
$$
\begin{equation}\label{c4}
\frac{1}{\vec q_1^{\:2}}\int\frac{d^2k}{\vec k^{\:2}(\vec k - \vec n)^2}
\ln \vec k^{\:2} \theta\left( \vec k^{\:2} - (\vec k - \vec n)^2
\right) = 0~,
\end{equation}
where $k$ is a dimensionless vector and $n$ an arbitrary vector with
$\vec n^{\:2} = 1$. Then one can make two changes of integration variables
\begin{equation}\label{c5}
\vec k \rightarrow \frac{\vec k}{\vec k^{\:2}},\ \ \ \vec k \rightarrow
- (\vec k - \vec n)
\end{equation}
to obtain
\begin{equation}\label{c6}
I = \frac{1}{\vec q_1^{\:2}}\int\frac{d^2k}{\vec k^{\:2}}\ln(\vec k - \vec n)^2
\theta(1 - \vec k^{\:2}) = \frac{1}{\vec q_1^{\:2}}\int_0^{2\pi}d\phi\int_0^1
\frac{dk}{k}\ln(1 + k^2 - 2k \cos\phi).
\end{equation}
As last step the integration by parts over $k$ gives an expression which
can be easily integrated over $\phi$ with zero result:
\begin{equation}\label{c7}
I = \frac{2\pi}{\vec q_1^{\:2}}\int_0^1\frac{dk}{k}\ln k\biggl[ \frac
{(1 - k^2)}{2\pi}\int_0^{2\pi}\frac{d\phi}{(1 + k^2 - 2k \cos\phi)} -1 \biggr]
= 0.
\end{equation}

\section{Discussion}
\setcounter{equation}{0}

In this paper we have obtained the NLA non-forward gluon impact
factors with any color structure in the $t$-channel in QCD with massive
quarks at arbitrary space-time dimension $D = 4 + 2\epsilon$.
Then we have used the integral representation in the case of the octet
impact factors to check the second bootstrap condition~\cite{12} and have 
found that it is satisfied. As it was mentioned above, this fact is very 
important from the theoretical point of view for the BFKL approach and 
demonstrates in very clear way the compatibility of this approach with the 
$s$-channel unitarity in the NLA. After this check we have carried
out integrations in the form of an expansion in $\epsilon$ in the
expression for the impact factor (\ref{28}) in the important 
case of QCD with $n_f$ massless quark flavors. For this 
last case the forward impact factor can be calculated exactly as a function 
of $\epsilon$. As an example we have presented in Eq.~(\ref{710}) the
singlet color helicity conserving impact factor. It was already obtained
in Ref.~\cite{17} as an expansion in $\epsilon$ with the accuracy up to terms
finite at $\epsilon \rightarrow 0$. We notice that, expanding our
exact result (\ref{710}), leads to an expression which is in
agreement with the result of Ref.~\cite{17} when taking into account the
differences in the definitions of impact factors. 

Let us note, that throughout this paper we have 
used the unrenormalized coupling constant $g$, regularizing both ultraviolet
and infrared divergences by the same parameter $\epsilon$. The
ultraviolet divergences are contained only in the one-gluon contribution
and come from the one-loop correction to the gluon-gluon-Reggeon effective 
interaction vertex (see Section 3). This vertex was defined in
Refs.~\cite{6},~\cite{14} and~\cite{15} in such way that the
ultraviolet divergences can be removed by simple charge renormalization in
the $\overline{MS}$ scheme
\begin{equation}\label{81}
g = g(\mu)\mu^{-\epsilon}\left[ 1 + \left( \frac{11}{3} - \frac{2}{3}\frac{n_f}
{N} \right)\frac{g^2(\mu)N\Gamma(1 - \epsilon)}{2\epsilon(4\pi)^{2 + \epsilon}}
\right]~.
\end{equation}
After this renormalization has been performed there are still divergences
of the infrared kind in the gluon impact factors because the gluon is not 
a colorless object. Note, that the definition of impact factors given 
in~\cite{12}  and used in this paper guarantees the infrared safety of 
the impact factors for colorless particles ~\cite{FM99}. 

\vskip 1.5cm \underline {Acknowledgment}: One of us (M.I.K.) thanks the
Dipartimento di Fisica della Universit\`a della Calabria and the Istituto
Nazionale di Fisica Nucleare - Gruppo collegato di Cosenza for their warm
hospitality while part of this work was done.

\appendix

\section{Appendix A}

In this section we calculate the integrals $K_1$ (see Eq.~(\ref{521})), 
$K_2$ - $K_4$ and $K_{4\perp\perp}^{\mu\nu}$ (see Eq.~(\ref{528})),
appearing in the expressions (\ref{75}) and  (\ref{76}) for quark-antiquark 
and two-gluon contributions to the gluon NLA non-forward impact factors 
(\ref{28}) in completely massless QCD. Let us firstly consider $K_1$. For 
the non-forward case we can present $K_1$ as follows:
$$
K_1 = \left( \vec q^{\:2} \right)^{\epsilon}{\widetilde K}_1~,
$$
\begin{equation}\label{A1}
{\widetilde K}_1 = \frac{(4\pi)^{2 + \epsilon}\Gamma(1 + 2\epsilon)}
{4\Gamma(1 - \epsilon)\Gamma^2(1 + \epsilon)}\int\frac{d^{D-2}k}{(2\pi)^{D-1}}
\ln\left( \frac{1}{\vec k^{\:2}} \right)\frac{1}{(\vec k - \vec k_1)^2(\vec
k - \vec k_2)^2}~,
\end{equation}
with
\begin{equation}\label{A2}
\vec k_1 = \frac{\vec q_1^{\:\prime}}{|\vec q|}~,\ \ \ \ \vec k_2 = 
\frac{\vec q_1}{|\vec q|}~,\ \ \ \ (\vec k_1 - \vec k_2)^2 = 1~,\ \ \ \ 
\vec q \neq 0~.
\end{equation}
We need for ${\widetilde K}_1$ an expression having accuracy up to terms of
the type $\epsilon^0$ and being exact in the regions $\vec k_1 \rightarrow 0$,
$\vec k_2 \rightarrow 0$, according to the discussion after 
Eq.~(\ref{76}). With the help of the equality
\begin{equation}\label{A3}
\ln\left( \frac{1}{\vec k^{\:2}} \right) = \frac{d}{d\alpha}\left( \frac{1}
{\vec k^{\:2}} \right)^{\alpha}\biggl|_{\alpha = 0}
\end{equation}
and of the generalized Feynman parametrization (see, for instance,
Ref.~\cite{5}) the integration over $\vec k$ gives
$$
{\widetilde K}_1 = \frac{\Gamma(1 + 2\epsilon)}{2\Gamma(1 - \epsilon)\Gamma^2
(1 + \epsilon)}\int_0^1dz\biggl( \frac{\Gamma(1 - \epsilon + \alpha)}{\Gamma
(1 + \alpha)}\alpha
$$
\begin{equation}\label{A4}
\int_0^1\frac{dxx^{\alpha - 1}(1 - x)^{\epsilon - \alpha}}{\left[ x\left( (1 -
z)\vec k_1^{\:2} + z\vec k_2^{\:2} \right) + (1 - x)z(1 - z) \right]^{1 -
\epsilon + \alpha}} \biggr)_{\alpha = 0}^{\prime}~.
\end{equation}
It is not difficult now to obtain a linear term in $\alpha$ in the expression
inside the large brackets of the last relation in order to perform the 
differentiation. Then the integral ${\widetilde K}_1$ takes the form
$$
{\widetilde K}_1 = \frac{1}{\epsilon}\biggl( \frac{1}{\epsilon} + 2\psi
(1 + 2\epsilon) - 2\psi(1 + \epsilon) + \psi(1 - \epsilon) - \psi(1) \biggr)
$$
$$
+ \frac{\Gamma(1 + 2\epsilon)}{2\Gamma^2(1 + \epsilon)}\biggl[ \int_0^1\int
_0^1dz\,dx\frac{\epsilon\ln x}{(1 - x)}\frac{(1 - x)^{\epsilon}}{\left[ x\left(
(1 - z)\vec k_1^{\:2} + z\vec k_2^{\:2} \right) + (1 - x)z(1 - z) \right]
^{1 - \epsilon}}
$$
\begin{equation}\label{A5}
+ \int_0^1\int_0^1 dz\,dx\ln x\frac{(1 - x)^{\epsilon}(1 - \epsilon)\left( (1 -
z)\vec k_1^{\:2} + z\vec k_2^{\:2} - z(1 - z) \right)}{\left[ x\left( (1 - z)
\vec k_1^{\:2} + z\vec k_2^{\:2} \right) + (1 - x)z(1 - z) \right]^{2 -
\epsilon}} \biggr]~.
\end{equation}
This expression is still exact and holds for any $\epsilon$. We should now 
perform the expansion in $\epsilon$ to carry out the integrations in 
Eq.~(\ref{A5}) in an approximate form. As for the first integral in the
R.H.S. of this equation, we find that it is order $O(\epsilon)$ and we 
neglect it in our calculation. In order to evaluate the second integral, we
divide the region of integration over $x$ in two parts in the following way:
\begin{equation}\label{A6}
1)\ \ 0 < x < \delta~,\ \ \ \ \ 2)\ \ \delta < x < 1~,
\;\;\;\;\;\;\;\delta
\rightarrow 0~.
\end{equation}
In the first region the integration becomes simpler and gives without any
difficulties the following contribution to the square brackets in the
R.H.S. of Eq.~(\ref{A5}):
\begin{equation}\label{A7}
- \frac{1}{\epsilon^2}\left( \left( \vec k_1^{\:2} \right)^{\epsilon} + \left(
\vec k_2^{\:2} \right)^{\epsilon} \right) + \ln^2 \delta ~,
\end{equation}
with accuracy up to $\epsilon^0$ type terms. In the second region, we 
can evidently put $\epsilon = 0$, because the integration is convergent. 
The contribution of this integration region, analogous to that of Eq.~(\ref{A7}), is
$$
- \ln^2 \delta - \ln \delta \ln(\delta\: \vec k_1^{\:2}\vec k_2^{\:2}) + \int
_0^1dz\int_{\delta}^1\frac{dx}{x}\frac{1}{\left[ x\left( (1 - z)\vec k_1^{\:2}
+ z\vec k_2^{\:2} \right) + (1 - x)z(1 - z) \right]}
$$
\begin{equation}\label{A8}
= - \ln^2 \delta  + \ln \vec k_1^{\:2} \ln \vec k_2^{\:2}~.
\end{equation}
As it must be, the 
dependence on $\delta$ is cancelled in the sum of the contributions
(\ref{A7}) and (\ref{A8}), so that the limit $\delta \rightarrow 0$ does
exist for this sum. Expanding with the required accuracy the first term
and the coefficient in front of the square brackets in Eq.~(\ref{A5}), we
get
$$
 \frac{1}{\epsilon}\biggl( \frac{1}{\epsilon} + 2\psi(1 + 2\epsilon) - 
2\psi(1 + \epsilon) + \psi(1 - \epsilon)- \psi(1) \biggr) = 
\frac{1}{\epsilon^2} + \psi^{\prime}(1)~,
$$
\begin{equation}\label{A10}
\frac{\Gamma(1 + 2\epsilon)}{2\Gamma^2(1 + \epsilon)} = \frac{1}{2}\left(
1 + \epsilon^2\psi^{\prime}(1) \right)~.
\end{equation}
Using now Eqs.~(\ref{A7}) and (\ref{A8}), for non-zero $\vec k_1$ or $\vec k_2$
(these vectors cannot be both zero at the same time, because $\vec q \neq
0$, according the definitions (\ref{A2})) we obtain 
\begin{equation}\label{A11}
{\widetilde K}_1 = - \frac{1}{2\epsilon}\ln\left( \vec k_1^{\:2}\vec k_2^{\:2}
\right) - \frac{1}{4}\ln^2\left( \frac{\vec k_1^{\:2}}{\vec k_2^{\:2}} \right).
\end{equation}
We should now include the correct asymptotics in the limits of small
$\vec k_1$ or $\vec k_2$.
It is easy to check that the correct asymptotics of ${\widetilde K}_1$ in 
the limit, for example, $\vec k_1 \rightarrow 0$, is given by the expression
$$
{\widetilde K}_1\left( \vec k_1 \rightarrow 0 \right) = \frac{(4\pi)^{2 +
\epsilon}\Gamma(1 + 2\epsilon)}{4\Gamma(1 - \epsilon)\Gamma^2(1 + \epsilon)}
\biggl[ \left( \vec k_1^{\:2} \right)^{\epsilon}\int\frac{d^{D-2}k}{(2\pi)^
{D-1}}\ln\left( \frac{1}{\vec k^{\:2}} \right)\frac{1}{(\vec k - \vec n)^2}
$$
\begin{equation}\label{A12}
+ \int\frac{d^{D-2}k}{(2\pi)^{D-1}}\ln\left( \frac{1}{\vec k^{\:2}} \right)
\frac{1}{\vec k^{\:2}(\vec k - \vec n)^2} \biggr]~,
\end{equation}
where $\vec n$ is an arbitrary vector with $\vec n^{\:2} = 1$, the omitted 
terms being at least of order $|\vec k_1|$  in the region under
consideration. The integrals in the R.H.S. of Eq.~(\ref{A12}) can be easily 
calculated and, using also the symmetry of ${\widetilde K}_1$ under the 
replacement $\vec k_1 \leftrightarrow \vec k_2$, we arrive at the
conclusion that the function
\begin{equation}\label{A13}
\frac{1}{2\epsilon^2}\left( 2 - \left( \vec k_1^{\:2} \right)^{\epsilon} -
\left( \vec k_2^{\:2} \right)^{\epsilon} \right) + \frac{1}{\epsilon}\biggl(
\psi(1 + 2\epsilon) - \psi(1 + \epsilon) + \psi(1 - \epsilon) - \psi(1)
\biggr)
\end{equation}
contains the correct asymptotics of ${\widetilde K}_1$ in both limits 
$\vec k_1,\ \vec k_2 \rightarrow 0$. To avoid double counting when
including this asymptotics into the intermediate result (\ref{A11}), we 
should add the function (\ref{A13}) in its exact form there and subtract it
in the expression obtained by expanding in $\epsilon$ up to terms non-vanishing in
the limit $\epsilon \rightarrow 0$. The final result for $K_1$
defined by Eqs.~(\ref{A1}) and (\ref{A2}) takes the form
\begin{equation}\label{A14}
K_1 = \frac{1}{2}\left( \vec q^{\:2} \right)^{\epsilon}\biggl[
\frac{1}{\epsilon^2}\biggl( 2 - \left( \frac{\vec q_1^{\:\prime\:2}}{\vec q^{\:2}}
\right)^{\epsilon} - \left( \frac{\vec q_1^{\:2}}{\vec q^{\:2}} \right)
^{\epsilon} \biggr) + 4\psi^{\prime\prime}(1)\epsilon + \ln\left( \frac{\vec
q_1^{\:\prime\:2}}{\vec q^{\:2}} \right)\ln\left( \frac{\vec q_1^{\:2}}{\vec q
^{\:2}} \right) \biggr]~,
\end{equation}
where we did not expand in $\epsilon$ the common factor
$\left(\vec q^{\:2} \right)^{\epsilon}$, because it would produce
an additional complication, although such an expansion is possible
in Eq.~(\ref{A14}).

Our next step is the calculation of $K_2$, defined in Eq.~(\ref{528}). It 
can be written in the following form:
\begin{equation}\label{A15}
K_2 = \biggl( - \frac{1}{2\epsilon} + \psi(1 + 2\epsilon) - \psi(1) \biggr)
\biggl( \left( \vec q_1^{\:2} \right)^{\epsilon} + \left( 
\vec q_1^{\:\prime\:2} \right)^{\epsilon} \biggr) + 2\epsilon 
\int_0^{\infty}dz\:\frac{\ln z}{(1 + z)^{1 + 2\epsilon}}\frac
{\left( \vec q(\vec q_1 + z\vec q_1^{\:\prime}) \right)}{\left( (\vec q_1 +
z\vec q_1^{\:\prime})^2 \right)^{1 - \epsilon}}~.
\end{equation}
In the region $q_1,\ q_1^{\prime} \neq 0$ (here and everywhere below
$k = |\vec k|$), when we can restrict ourselves to consider $K_2$ with
accuracy up to terms linear in $\epsilon$ (see Eq.~(\ref{76})), the
integration in Eq.~(\ref{A15}) is performed in a quite straightforward
way and the final result takes the form
\begin{equation}\label{A16}
K_2 = - \frac{1}{\epsilon} - \ln(q_1 q_1^{\prime}) + 4\epsilon\psi^{\prime}(1)
- 2\epsilon\ln q_1 \ln q_1^{\prime} - \epsilon\theta^2,
\end{equation}
with
\begin{equation}\label{Aa1}
\cos\theta = \frac{(\vec q_1\vec q_1^{\:\prime})}{q_1q_1^{\prime}},\ \ \
0 < \theta < \pi.
\end{equation}
Unfortunately, the correct asymptotics of the R.H.S. of Eq.~(\ref{A15}) in
the limits $q_1 \rightarrow 0$ and $q_1^{\prime} \rightarrow 0$
cannot be expressed in terms of elementary functions only, so that it seems
to be better to leave the integral $K_2$ in the exact form of Eq.~(\ref{A15}),
which is convenient for subsequent applications of the gluon impact factors
(\ref{28}).

The remaining integrations of $K_3$, $K_4$ and $K_{4\perp\perp}^{\mu\nu}$
(\ref{528}) are easily performed and the results we obtain for the
non-forward case $q \neq 0$, valid also in the regions of small
$q_1$ or $q_1^{\prime}$, are:
\begin{equation}\label{A17}
K_3 = 1 - 2\epsilon + \epsilon\ln(q_1 q_1^{\prime}) + \frac{\epsilon}{q^2}
\left( (q_1^2 - q_1^{\prime\:2})\ln\left( \frac{q_1}{q_1^{\prime}} \right)
+ 2q_1q_1^{\prime}\theta \sin\theta \right) + 2\epsilon^2\left( 2 - 2\ln q
+ \ln^2 q \right)~,
\end{equation}
$$
K_4 = \frac{1}{6} - \frac{5\epsilon}{18} + \frac{\epsilon}{6}\ln(q_1 q_1
^{\prime}) - \frac{2\epsilon q_1q_1^{\prime}}{3q^4}\left( 2q_1q_1^{\prime}
\sin^2\theta - q^2\cos\theta \right) + \frac{\epsilon(q_1^2 - q_1^{\prime\:2})}
{6q^6}\left( 8q_1^2q_1^{\prime\:2}\sin^2\theta \right .
$$
$$
\left. - 2q_1q_1^{\prime}q^2 \cos\theta
+ q^4 \right)\ln\left( \frac{q_1}{q_1^{\prime}} \right) + \frac{2\epsilon
q_1^2q_1^{\prime\:2}}{3q^6}\left( 4q_1q_1^{\prime}\sin^2\theta - 3q^2\cos\theta
\right)\theta \sin\theta 
$$
\begin{equation}\label{A18}
+ \frac{\epsilon^2}{54}\left( 19 - 30\ln q +
18\ln^2 q \right)~,
\end{equation}
$$
K_{4\perp\perp}^{\mu\nu} = \frac{\left( q_{1\perp}^{\prime\:\mu}q_{1\perp}
^{\prime\:\nu} + q_{1\perp}^{\mu}q_{1\perp}^{\nu} \right)}{2q_{\perp}^2}\biggl[
- 1 + \frac{(q_1^2 - q_1^{\prime\:2})}{q^2}\ln\left( \frac{q_1}{q_1^{\prime}}
\right) + \left( 2q_1q_1^{\prime}\sin^2\theta - q^2\cos\theta \right)\frac
{\theta}{q^2\sin\theta} \biggr]
$$
$$
+ \frac{\left( q_{1\perp}^{\prime\:\mu}q_{1\perp}^{\prime\:\nu} - q_{1\perp}^{\mu}
q_{1\perp}^{\nu} \right)}{2q_{\perp}^2}\biggl[ - \frac{2(q_1^2 - q_1^{\prime
2})}{q^2} + \ln\left( \frac{q_1}{q_1^{\prime}} \right) + \frac{2q_1q_1
^{\prime}}{q^4}\left( 3q^2\cos\theta - 4q_1q_1^{\prime}\sin^2\theta \right)\ln
\left( \frac{q_1}{q_1^{\prime}} \right)
$$
$$
+ \left( 4q_1q_1^{\prime}\sin^2\theta - q^2\cos\theta \right)\frac{(q_1^2 -
q_1^{\prime\:2})\theta}{q^4\sin\theta} \biggr] - \frac{q_{\perp}^{\mu}q_{\perp}
^{\nu}}{q_{\perp}^2}\biggl[ \frac{1}{3} + \frac{\epsilon}{18}\left( 5 - 6\ln q
\right) 
$$
$$
+ \frac{q_1q_1^{\prime}}{q^4}\left( 3q^2\cos\theta - 4q_1q_1^{\prime}
\sin^2\theta \right) + \frac{2q_1q_1^{\prime}}{q^6}(q_1^2 - q_1
^{\prime\:2})\left( 2q_1q_1^{\prime}\sin^2\theta - q^2\cos\theta \right)\ln\left(
\frac{q_1}{q_1^{\prime}} \right)
$$ 
\begin{equation}\label{A19}
\biggl.
+ \left( q^4 - 2(q_1^2 - q_1^{\prime\:2})^2\sin
^2\theta \right)\frac{q_1q_1^{\prime}\theta}{q^6\sin\theta} \biggr]~.
\end{equation}

\end{document}